\documentclass{aa}  

\usepackage{graphicx}
\usepackage{txfonts}

\usepackage{multirow}

\usepackage{hyperref}
\newcommand{\caesar}{\textsc{caesar}}

\newcommand{\hii}{H\textsc{ii}}

\defcitealias{Anderson2014}{A19}

\begin{document}

   \title{The SARAO MeerKAT Galactic Plane Survey extended source catalogue}

   \author{C. Bordiu \inst{1}
          \and
          S. Riggi \inst{1} \and
          F. Bufano \inst{1} \and
          F. Cavallaro \inst{1} \and
          T. Cecconello\inst{1,2} \and
          F. Camilo \inst{3} \and
          G.Umana \inst{1} \and
          W. D. Cotton \inst{4} \and
          M. A. Thompson \inst{5} \and
          M. Bietenholz \inst{6,7} \and
          S. Goedhart \inst{3,8} \and
          L. D. Anderson \inst{9,10,11} \and
          C. S. Buemi \inst{1} \and
          J. O. Chibueze \inst{12,13} \and
          A. Ingallinera \inst{1} \and
          P. Leto \inst{1} \and
          S. Loru \inst{1} \and
          M. Mutale \inst{5} \and
          A. Rigby \inst{5} \and
          C. Trigilio \inst{1} \and
          G. M. Williams \inst{5,14}
          }

   \institute{INAF-Osservatorio Astrofisico di Catania, Via Santa Sofia 78, 95123 Catania, Italy\\
              \email{cristobal.bordiu@inaf.it, simone.riggi@inaf.it}
         \and
             Department of Electrical, Electronic and Computer Engineering, University of Catania, Viale Andrea Doria 6, 95125 Catania, Italy
        \and 
            South African Radio Astronomy Observatory, 2 Fir Street, Observatory 7925, South Africa
        \and
            National Radio Astronomy Observatory, 520 Edgemont Road, Charlottesville, VA 22903, USA
        \and
            School of Physics and Astronomy, University of Leeds, Leeds LS2 9JT, UK
        \and
            SARAO/Hartebeesthoek Radio Astronomy Observatory, PO Box 443, Krugersdorp 1740, South Africa
        \and
            Department of Physics and Astronomy, York University, Toronto, M3J 1P3, Ontario, Canada
        \and
            SKA Observatory, 2 Fir Street, Observatory 7925, South Africa
        \and 
            Department of Physics and Astronomy, West Virginia University, Morgantown, WV 26506, USA
        \and 
            Adjunct Astronomer at the Green Bank Observatory, P.O. Box 2, Green Bank, WV 24944, USA
        \and
            Center for Gravitational Waves and Cosmology, West Virginia
University, Chestnut Ridge Research Building, Morgantown, WV 26505, USA
        \and
            Department of Mathematical Sciences, University of South Africa, Cnr Christian de Wet Rd and Pioneer Avenue, Florida Park, 1709, Roodepoort, South Africa
        \and
            Department of Physics and Astronomy, Faculty of Physical Sciences, University of Nigeria, Carver Building, 1 University Road, Nsukka 410001, Nigeria
        \and
            Department of Physics, Aberystwyth University, Ceredigion, Cymru, SY23 3BZ, UK}

   \date{Received xxx; accepted yyy}

 
  \abstract
  {We present a catalogue of extended radio sources from the SARAO MeerKAT Galactic Plane Survey (SMGPS). Compiled from 56 survey tiles and covering approximately 500 deg$^2$ across the first, third, and fourth Galactic quadrants, the catalogue includes 16534 extended and diffuse sources with areas larger than 5 synthesised beams. Of them, 3891 (24\% of the total) are confidently associated with known Galactic radio-emitting objects in the literature, such as \hii{} regions, supernova remnants, planetary nebulae, luminous blue variables, and Wolf-Rayet stars. A significant fraction of the remaining sources, 5462 (33\%), are candidate extragalactic sources, while 7181 (43\%) remain unclassified. Isolated radio filaments are excluded from the catalogue. The diversity of extended sources underscores MeerKAT's contribution to the completeness of censuses of Galactic radio emitters, and its potential for new scientific discoveries. For the catalogued sources, we derived basic positional and morphological parameters, as well as flux density estimates, using standard aperture photometry. This paper describes the methods followed to generate the catalogue from the original SMGPS tiles, detailing the source extraction, characterisation, and crossmatching procedures. Additionally, we analyse the statistical properties of the catalogued populations.}


   \maketitle
%

\section{Introduction}

Deep radio continuum surveys provide answers to a number of fundamental questions in astrophysics. In particular, radio observations of the Galactic plane advance our understanding of the structure and evolution of the Milky Way, allowing us to peer through the obscuring dust that blocks shorter wavelengths and thus revealing otherwise hidden sources and phenomena.

Since the early days of radio astronomy, numerous radio continuum Galactic plane surveys have been carried out at different facilities, both in the northern hemisphere --- for example, the Multi-Array Galactic Plane Imaging Survey (MAGPIS; \citealt{magpis1}), the Very Large Array (VLA) Galactic Plane Survey (VGPS; \citealt{vgps}), the Coordinated Radio and Infrared Survey for High-Mass Star Formation (CORNISH; \citealt{cornish}), and the HI/OH/Recombination Line Survey of the Milky Way (THOR; \citealt{thor}) --- and the southern hemisphere --- for example, the Molonglo Galactic Plane Survey (MGPS; \citealt{mgps}) and CORNISH-South (\citealt{irabor23}) --- charting extensive portions of the sky and uncovering thousands of new radio-emitting sources all across the Galactic plane. For decades, the  sensitivity and angular resolution of world-class radio interferometers such as the VLA or ATCA have continued to improve, and they, sometimes in combination with single-dish telescopes \citep{vgps, sgps, magpis2}, have provided the most complete and accurate surveys to date, principally in the low-frequency regime (up to a few gigahertz), where emission originates from both thermal (free-free emission from hot plasma) and non-thermal processes (e.g. synchrotron radiation from relativistic electrons). Radio continuum surveys, together with complementary multi-wavelength data, have greatly enhanced our understanding of the Galaxy and its constituents, shedding light on some of the hottest astrophysical topics: the physics of the Galactic centre \citep{Yusef2004,Barkov19}, the structure of the ionised interstellar medium \citep{Thompson2015, Umana2021}, and the stellar life cycle --- from star formation to the elusive final evolutionary stages of stars across the entire mass spectrum \citep{Umana2015, Wang2018, Bojicic2021, Ball2023}.

All surveys are forced to compromise between survey area, depth, and angular resolution. For instance, surveys such as VGPS \citep{vgps} and SGPS \citep{sgps} almost completely cover the first and fourth Galactic quadrants with $\sim$1 arcmin resolution  and millijansky beam$^{-1}$ sensitivity, whereas higher-resolution surveys such as CORNISH ($\sim$1.5 arcsec; \citealt{cornish}) and MAGPIS ($\sim$6 arcsec; \citealt{magpis1}) have somewhat more limited spatial extents. Moreover, additional constraints in terms of $uv$ coverage significantly impact imaging fidelity and dynamic range, affecting the detection of extended low-surface-brightness radio sources. 

These limitations have driven significant technological upgrades to existing observing facilities and inspired the development of new ones with dense instantaneous $uv$ coverage, pushing the technical boundaries of radio continuum surveys \citep{Norris2013}. In this context, instruments such as ASKAP \citep{Johnston2008}, MeerKAT \citep{Jonas2009, Jonas2016}, LOFAR \citep{vanHaarlem2013}, and MWA \citep{Lonsdale09}, as precursors and pathfinders of the Square Kilometre Array (SKA), enable fast and deep large-area continuum surveys with unparalleled detail. These facilities can provide nearly complete censuses of Galactic radio emitters, enabling unbiased population studies and unlocking unexpected scientific discoveries that profoundly impact many astrophysical fields. Successful pilot programmes such as the observations of the SCORPIO (Stellar Continuum Originating from Radio Physics In Ourgalaxy) field with ASKAP \citep{Umana2021,Riggi2021} and the Galactic Centre with MeerKAT \citep{Heywood2022, Yusef2022} offer a glimpse of their immense potential for Galactic science.

The MeerKAT telescope array stands as one of the spearheads of the upcoming revolution. Located in Northern Cape, South Africa, it comprises 64 offset Gregorian dishes, 13.5 m in diameter, with a total collecting area of $\sim$9000 m$^2$. The South African Radio Astronomy Observatory (SARAO) MeerKAT Galactic Plane survey (hereafter SMGPS) was conducted between 2018 July 21 and 2020 March 14 with 60 antennas, imaging a large fraction of the first, third, and fourth Galactic quadrants ($l=2\degr-60\degr, 252\degr-358\degr$, $b=\pm1.5\degr$) in the L band (886--1678 MHz).

The SMGPS is currently the largest, most sensitive, and highest-angular-resolution L-band Galactic plane survey to date. In this paper, we present a catalogue of the extended radio sources in the SMGPS. For the catalogued sources, we derived basic global parameters (position and flux density) and studied their statistical distribution. When possible, we also provide a tentative classification of the detected structures by crossmatching with catalogues of known Galactic sources. In particular, our crossmatch focuses on the different stages of the stellar life cycle: \hii{} regions, probing areas of active star formation; planetary nebulae (PNe) around evolved low- to intermediate-mass stars; circumstellar ionised structures surrounding evolved high-mass stars such as luminous blue variables (LBVs) and Wolf-Rayet (WR) stars; and supernova remnants (SNRs). All of these sources are radio emitters at gigahertz frequencies and typically appear as extended structures in high-resolution observations \citep{Gudel2002,Dubner2015}. An in-depth study of these extended sources and their physical properties is beyond the scope of this paper but will be the subject of a series of publications focusing on SNRs \citep{Loru24}, PNe (Ingallinera et al., in prep.), LBVs (Umana et al., in prep.), and WR stars (Buemi et al., in prep.).

The paper is organised as follows: In Sect. \ref{sec:observations} we briefly describe the SMGPS observations and data, as well as other ancillary datasets used in this work. In Sect. \ref{sec:catalog} we detail the source extraction, characterisation, and crossmatching procedures employed to generate the catalogue. Section \ref{sec:analysis} presents a global analysis of the resulting catalogue, focusing on source distribution and statistical properties. Finally, in Sect. \ref{sec:summary} we summarise the main outcomes of this work and its added value in the context of the forthcoming SKA.

\section{Radio observations}
\label{sec:observations}

\subsection{SMGPS: Observations and available data}
\label{sec:SMGPS}

 A full technical description of the radio observations and the data reduction procedures is provided in \cite{paper1}. The SMGPS comprises a set of 56 overlapping primary beam-corrected mosaics, each one covering a patch (hereafter \lq tile\rq) of the sky of area $\sim$3$^{\circ}$ $\times$ 3$^{\circ}$. For each tile, the standard data products released in DR1 consist of: (1) a frequency plane cube, with a fitted broadband flux density plane at a reference frequency of 1359.7 MHz, a spectral index plane, and 14 individual frequency planes; (2) a fitted-parameter cube, including the same first two planes (broadband flux density and spectral index) plus additional channels for the error estimate of the broadband flux fit, and the least squares and $\chi^2$ of the spectral index fit; and (3) a zeroth-moment integrated intensity map, with a characteristic effective frequency and bandwidth of 1293 and 672 MHz, respectively. The fitted broadband flux density plane in the data cubes excludes the mosaic regions at high Galactic latitudes that were only observed at low frequencies due to the frequency-dependent primary beam. In contrast, the zeroth-moment maps, which are computed as a weighted sum of the available pixels across the band, provide more complete sky coverage (see  \citealt{paper1} for further details). Therefore, in this work we employed the zeroth-moment maps (`maps' hereafter) as the main reference for source finding, and the fitted flux density plane for flux density measurements. The remaining individual frequency planes are instead reserved for follow-up spectral studies of different source populations.

The final mosaics have a frequency-independent circular synthesised beam of 8"$\times$8", and a background rms noise, measured in areas far from the Galactic plane, of the order of 10--20 $\mu$Jy beam$^{-1}$. MeerKAT has a dense short baseline coverage but lacks zero-spacing information. Hence, the imaging of extended structures is in principle limited by the minimum baseline length of 29 m, which translates into a theoretical largest angular scale (LAS) ranging from $\sim$21 to $\sim$40 arcmin across the L-band frequency range. At the representative frequency of the maps, the LAS is $\sim$27 arcmin. However, this value is just an approximate upper limit on the source scale to which the instrument is sensitive. In practice, the shallow \textsc{clean} used for the SMGPS images (single-scale, with a maximum of 250,000 components per pointing, and a deconvolution depth of 100--200 $\mu$Jy beam$^{-1}$; \citealt{paper1}) further limits the recovery of flux density. This implies
 that scales significantly smaller than 27 arcmin may not be fully recovered.

Both positional and flux density systematic uncertainties have been studied using compact sources  in \cite{paper1}. The astrometry of the survey is found to be accurate to a level of $\sim$0.5 arcsec, and therefore the impact on extended sources is negligible. The systematic uncertainty in the flux density scale is 5\% as determined for point sources through comparison with existing surveys, and will be used as the reference value in the remainder of the paper. 

\begin{table*}
\centering%
\scriptsize%
\caption{SMGPS parameters compared to previous radio continuum surveys covering the Galactic plane in the frequency range 0.8--6.0 GHz.}
\begin{tabular}{lllllcccccl}
\hline%
\hline%
Survey & Instr. & $l$ Coverage & $b$ Coverage & Quadrant & Freq. & Bandwidth & FWHM & rms & LAS & Ref.\\%
& & (deg) & (deg) & & (GHz) & (MHz) & (arcsec) & ($\mu$Jy/beam) & (arcmin) & \\%
\hline%
\multirow{2}{*}{SMGPS} & \multirow{2}{*}{MeerKAT} & 2$<l<$60 & $|b|<$1.5 & I & \multirow{2}{*}{1.284} & \multirow{2}{*}{792} & \multirow{2}{*}{8} & \multirow{2}{*}{30} & \multirow{2}{*}{27} & \multirow{2}{*}{1}\\%
& & 252$<l<$358 & $|b|<$1.5 & III, IV\\%
\hline%
\multirow{3}{*}{VGPS} & \multirow{3}{*}{VLA+GBT} & 18$<l<$46 & $|b|<$1.3 & \multirow{3}{*}{I} & \multirow{3}{*}{1.4} & \multirow{3}{*}{1.866} & \multirow{3}{*}{60} & \multirow{3}{*}{--\tablefootmark{a}} & \multirow{3}{*}{--\tablefootmark{b}} & \multirow{3}{*}{2}\\%
& & 46$<l<$59 & $|b|<$1.9 & \\%
& & 59$<l<$67 & $|b|<$2.3 & \\%
\hline%
\multirow{2}{*}{MAGPIS (6 cm)} & \multirow{2}{*}{VLA} & 0$<l<$49.5 & $|b|<$1.0 & I & \multirow{2}{*}{5.0} & \multirow{2}{*}{50} & \multirow{2}{*}{6} & \multirow{2}{*}{179} & \multirow{2}{*}{--} & \multirow{2}{*}{3}\\%
& & 350<l<360 & $|b|<$0.4 & IV &\\%
\hline%
\multirow{4}{*}{MAGPIS (21 cm)} & VLA+Effelsberg & 5$<l<$48.5 & $|b|<$0.8 & I & 1.4 & 95 & 6 & 897 & --\tablefootmark{c} & 4\\%
\cline{2-11}%
& \multirow{3}{*}{VLA} & $-$20$<l<$120 & $|b|<$0.8 & I, IV & \multirow{3}{*}{1.4} & \multirow{3}{*}{--\tablefootmark{d}} & \multirow{3}{*}{6} & \multirow{3}{*}{897} & \multirow{3}{*}{--} & \multirow{3}{*}{3}\\%
& & $-$10$<l<$40 & $|b|<$1.7 & I, IV & & & & & & \\%
& & 100$<l<$105 & $|b|<$2.2$^{\circ}$ & I & & & & & & \\%
\hline%
CORNISH & VLA & 10$<l<$65 & $|b|<$1.1 & I & 5.0 & 25 & 1.5 & 400 & 2 & 5\\%
\hline%
CORNISH-South & ATCA & 295$<l<$350 & $|b|<$1 & IV & 5.5 & 2000 & 2.5 & 110 & -- & 6\\
\hline%
GLOSTAR & VLA & 28$<l<$36 & $|b|<$1 & I & 5.8 & --\tablefootmark{e} & 18 & 150 & 4 & 7\\%
\hline%
MGPS & MOST & 245$<l<$365 & $|b|<$10 & III, IV & 0.843 & 3 & 45$\times$45 csc$|\delta|$ & 1000 & 25 & 8\\%
\hline%
\multirow{2}{*}{SGPS} & \multirow{2}{*}{ATCA + Parkes} & 253$<l<$358 & $|b|<$1.5 & III, IV & \multirow{2}{*}{1.4} & \multirow{2}{*}{128} & 132 & \multirow{2}{*}{<1000} & \multirow{2}{*}{--\tablefootmark{f}} & \multirow{2}{*}{9}\\%
& & 5$<l<$20 & $|b|<$1.5 & I & & & 198\\%
\hline%
CGPS & DRAO Synth. Tel. & 74.2$<l<$147.3 & $-$3.6$<b<$5.6 & I, II & 1.42 & 35 & 60$\times$60 csc$|\delta|$ & 300 & 40 & 10\\%
\hline%
\multirow{2}{*}{THOR} & \multirow{2}{*}{VLA} & \multirow{2}{*}{14$<l<$67.4} & \multirow{2}{*}{$|b|<$1.25} & \multirow{2}{*}{I} & \multirow{2}{*}{1.42} & \multirow{2}{*}{128} & 18.1$\times$11.1 to & \multirow{2}{*}{300$-$1000} & \multirow{2}{*}{2} & \multirow{2}{*}{11}\\%
 & &  &  &  &  &  & 12.0$\times$11.6 &  & \\%
\hline%
\end{tabular}
\tablefoot{
\tablefoottext{a}{Noise level $\sim$0.3 K.}
\tablefoottext{b}{Without Green Bank Telescope data, the quoted LAS of VLA D-configuration is $\sim$16 arcmin.}
\tablefoottext{c}{Without Effelsberg data, the quoted LAS of VLA D-configuration is $\sim$16 arcmin.}
\tablefoottext{d}{Bandwidth variable across different observations, from 40 to 200 MHz.}
\tablefoottext{e}{Final mosaic obtained by integrating 8 sub-band images formed from two 1-GHz wide bands centred at 4.7 and 6.9 GHz.}
\tablefoottext{f}{Without Parkes data, the theoretical LAS of ATCA smallest baseline configuration is $\sim$23 arcmin.}
}
\tablebib{(1)~\citet{paper1};
(2) \citet{vgps}; (3) \citet{magpis1}; (4) \citet{magpis2};
(5) \citet{cornish}; (6) \citet{irabor23};
(7) \citet{glostar}; (8) \citet{mgps}; (9) \citet{sgps}; (10) \citet{cgps}; (11) \citet{thor}.
}
\label{tab:gpsurveys}
\end{table*}

\subsection{Comparison with other radio surveys}
\label{sec:gp-surveys}

The SMGPS angular resolution places it on par with some of the highest-resolution radio continuum surveys in the literature, with the added advantage of a superior sensitivity. Radio continuum surveys partially or fully covering the Galactic plane in the range of frequencies 0.8-6.0 GHz are reported in Table~\ref{tab:gpsurveys}. In the first Galactic quadrant, SMGPS observations are about one order of magnitude more sensitive than existing surveys carried out at similar frequencies, such as THOR \citep{thor} or MAGPIS \citep{magpis1,magpis2}, whereas angular resolution is, in general, comparable. The third and fourth Galactic quadrants are fully covered by the MGPS \citep{mgps} and SGPS \citep{sgps}  surveys, but with angular resolution ($\sim$1 arcmin) and rms ($\sim$1 mJy beam$^{-1}$) one order of magnitude worse than SMGPS. In view of these numbers, we expect the scientific potential of SMGPS to be primarily exploited in these quadrants, even if the improved sensitivity will certainly benefit first quadrant studies as well, enabling the detection of new faint and diffuse sources. 

Regarding $uv$ coverage, half of MeerKAT's baselines lie between 48 antennas within the array's inner core ($\sim$1 km in diameter), providing dense instantaneous sampling of the inner $uv$ plane, ideal for imaging extended emission. Additionally, the SMGPS observing strategy was designed to further enhance $uv$ coverage, revisiting each pointing multiple times over $\sim$10 h sessions (total on-source time $\sim$1 h; \citealt{paper1}). While this represents an improvement over other interferometric surveys with less dense samplings, the SMGPS still has limitations: first, the shallow deconvolution used in DR1 constrains the amount of flux density that is recovered from extended structures, with the effect fractionally worse for fainter sources (see Sect. \ref{subsec:biases} and Appendix \ref{sec:app-LAS}). Second, the absence of zero-baseline data results in extended bright features that are poorly sampled by the $uv$ coverage appearing surrounded by frequency-dependent negative bowls. This not only complicates flux density estimates but also artificially steepens spectral indices.

\subsection{Ancillary data and catalogues}
\label{sec:ancillary-catalogues}

To search for possible associations between MeerKAT sources and known Galactic objects, we considered the following source catalogues:
\begin{enumerate}
\item the Hong Kong/AAO/Strasbourg H-alpha (HASH) Planetary Nebula Database\footnote{\url{http://202.189.117.101:8999/gpne/dbMainPage.php}} \citep{Parker2016};
\item the WISE Catalogue of Galactic \hii{} regions v2.4\footnote{\url{http://astro.phys.wvu.edu/wise/}} (\citealt{Anderson2014}, last updated in 2019, henceforth \citetalias{Anderson2014});
\item the Catalogue of Galactic Supernova Remnants compiled by Green\footnote{\url{http://www.mrao.cam.ac.uk/surveys/snrs/}} \citep{Green2019}, along with newly identified SNRs from the THOR \citep{Anderson2017}, GLEAM \citep{Gleam2019}, and GLOSTAR \citep{2021A&A...651A..86D} surveys, including candidate and confirmed sources;
\item the Galactic Wolf-Rayet Star Catalogue\footnote{\url{http://pacrowther.staff.shef.ac.uk/WRcat/index.php}} \citep{Rosslowe2015};
\item the 2018 Census of Luminous Blue Variables in the Local Group \citep{Richardson2018}.
\end{enumerate}

\section{Catalogue of extended sources}
\label{sec:catalog}

The SMGPS tiles contain a wide variety of objects, whose emission is spread over a range of different angular scales and morphologies. This work is part of an extensive project to catalogue the complex emission within the SMGPS, structured into three main catalogues. Here, we focused exclusively on \lq extended sources\rq, defining an extended source as one that has an area of $>$5 synthesised beams. A complementary catalogue of \lq compact sources\rq\, containing point and resolved sources with areas $<$5 synthesised beams, $\sim$97\% complete at S/N>5, is presented in Mutale et al. (in prep.). The five-beam boundary is an arbitrary choice, driven by the different methodological approaches (see Sect. \ref{subsec:source-extraction}) and prioritising single component sources in the compact source catalogue. Additionally, a separate catalogue of radio filaments --- isolated thread-like radio structures resembling those found in the Galactic Centre \citep{Heywood2022} and requiring specialised extraction techniques --- is provided in \cite{Williams2024}. 

To build the extended source catalogue and throughout the remainder of this paper, we adopted the following terminology:
\begin{itemize}
\item Island (or source island): a group of at least four connected pixels with brightness above an aggregation threshold, surrounding a \lq seed\rq, pixel above a given detection threshold (5$\sigma$; \citealt{Han12,Han18,Riggi2023}).
\item Source: a generic astronomical object (e.g. irrespective of its morphology or astronomical nature) composed of one or more islands.
\item Compact sources: both unresolved (i.e. point-like) and resolved single islands that can be modelled well as a superposition of 2D Gaussians, having an area (expressed in number of synthesised beams) smaller than five beams.
\item Extended sources: single islands with well-defined edges and shapes departing from the circular Gaussian beam model, or with an area (in number of beams) larger than five beams. Multi-island sources (e.g. multi-island radio galaxies), including both compact and extended islands, are also categorised as extended sources. 
\item Diffuse sources: extended sources without well-defined edges, mostly large-scale structures.
\end{itemize}

In this work, we intentionally excluded isolated filamentary structures, namely those that do not form integral parts of extended sources (e.g. SNRs). Likewise, compact sources are only considered as part of masking operations, to exclude them from background and flux density estimates.

The method followed to build the catalogue is described in the following subsections. Details on how to access and download the catalogue are provided in the Data Availability section, and the full catalogue format is described in Appendix \ref{sec:app-cat-description}.

\subsection{Source extraction}
\label{subsec:source-extraction}

We employed a dual strategy to extract extended sources, integrating automated techniques with manual refinement after visual inspection. For each tile, we first extracted sources in an automated way using the algorithms implemented in the \caesar{} source finder \citep{Riggi2016,Riggi2019}. 
In particular, compact sources were extracted employing a standard flood-fill algorithm with the finder parameters reported in the appendix of \cite{Riggi2021}. Then, each extracted source was fit with multiple Gaussians. From the resulting source list, single-island sources showing a slightly extended morphology were extracted by applying the following selection criteria:

\begin{itemize}
\item number of fit components on island $>$3, or
\item number of beams on island $>$20, or
\item number of beams on island $>$10 and Gaussian source fit not converged or converged with poor quality ($\chi^{2}>$10).\end{itemize}

While this approach enabled the automatic extraction of numerous extended sources, the adopted criteria proved insufficient for most of the faint (i.e. below the 5$\sigma$ significance threshold that \caesar{} automatically imposes) extended or diffuse sources, which were frequently missed. We therefore considered other algorithms specifically designed for extended source detection, such as the saliency filter, the Wavelet Transform filter and the Chan-Vese active contour method described in \cite{Riggi2016}, and using parameters tuned with simulated data in \cite{Riggi2019}. These methods were able to recover some faint sources missed by the flood-fill algorithm, yet they generally yielded a high rate of spurious detections. Furthermore, these methods were unable to separate close bright extended sources or nested extended structures, which were typically extracted as a single source. In fact, all the employed algorithms, whether optimised for compact or extended sources, currently lack the ability to merge multiple disjointed islands into single sources. This is a critical challenge in certain scenarios, for example for radio galaxies that display separate core and jets. All these algorithmic issues were somewhat expected, given the inherent complexity of source extraction tasks, which are not straightforward even for human experts (\citealt{Riggi2019} and references therein).

To mitigate these limitations, we conducted a visual inspection of the tiles and the preliminary catalogue, which involved: (1) manually adding any missing extended, diffuse sources; (2) refining the segmentation of automatically extracted sources; and (3) including numerous small but clearly resolved sources --- generally with a bipolar or elongated morphology --- that failed to meet the \caesar{} selection criteria described above. The visual inspection of the catalogue required substantial effort and was conducted by multiple teams of professional astronomers in parallel, following common extraction guidelines to ensure consistency. To further reduce potential biases, automated validation procedures and cross-checks among different teams were performed. After completing the visual extraction and validation process, we implemented a minimum size threshold to ensure that only sources with areas larger than 5 synthesised beams were incorporated into the final catalogue --- corresponding to 94\% of the preliminary catalogue. The excluded 6\% comprises manually added sources that did not meet the size criterion and thus belong to the compact source catalogue (Mutale et al., in prep.).

For each tile, a curated source list was produced with the following specifications:
\begin{itemize}
\item Source islands are provided as DS9\footnote{\url{ds9.si.edu}} polygon regions delineating the overall source contour\footnote{In the case of diffuse sources, without well-defined contours, the provided segmentation is approximate.}. 
\item Each source is assigned a unique ID, and an `EXTENDED' or `DIFFUSE' label according to the aforementioned criteria;
\item Multiple disjointed source islands associated with the same physical object share the same ID and are consequently labelled `MULTIISLAND';
\item Sources can be nested, that is to say, they can contain a hierarchy of compact, extended, or diffuse sources.
\item Sources found at the edge of the tile are labelled `BORDER'. These sources may partially lie outside the SMGPS coverage area, resulting in incomplete photometry (the reported flux densities for such sources should be regarded as lower limits). In addition, source duplicates found in the overlapping area between contiguous tiles are removed from the catalogue.
\end{itemize}

\noindent Validated data products were finally put under data version control (git+DVC\footnote{\url{https://dvc.org/}}), for reproducibility of the analysis presented in Sect. \ref{sec:analysis}.

\subsection{Source crossmatching}
\label{subsec:source-crossmatch}
After extraction, each DS9 region was crossmatched with the ancillary catalogues of Galactic objects listed in Sect.~\ref{sec:ancillary-catalogues}, and a classification tag (SNR, HII, PN, LBV, or WR) was assigned in the case of a match. An additional `CONFIRMED' or `CANDIDATE' tag was given, depending on the status of the associated objects in their original catalogues (e.g. SNR candidates from THOR). 

The crossmatch procedure was conducted visually in order to minimise spurious associations: for stellar objects (i.e. LBVs, WRs, and PNe), we considered the reported catalogue positions to look for associated radio structures (such as circumstellar shells, bubbles, or nebulae). For \ion{H}{ii} regions and SNRs, which are potentially larger and more complex, we considered both the catalogue positions and angular sizes in the crossmatch. Still, the association of a segmented radio source with a catalogued source was not always clear, particularly in tiles close to the Galactic Centre, where confusion is higher. Ambiguous cases mostly fall under these categories:
\begin{enumerate}
\item Extended radio sources overlapping with the position of several \hii{} regions (hardly separable from each other), which were classified as \hii{} regions, and all matching object names were recorded (i.e. a single radio source is associated with multiple catalogued objects);
\item Extended radio sources clearly associated with catalogued radio-quiet \hii{} regions, which were classified as \hii{} regions and tagged as `CONFIRMED';
\item Extended radio sources with counterparts in more than one ancillary catalogue, which were given both tags, so that  multiple associations are kept in the final catalogue (e.g. an extended source coincident with both a known SNR and a known \hii{} region);
\end{enumerate}

\begin{figure*}
\centering%
\includegraphics[width=1.0\textwidth]{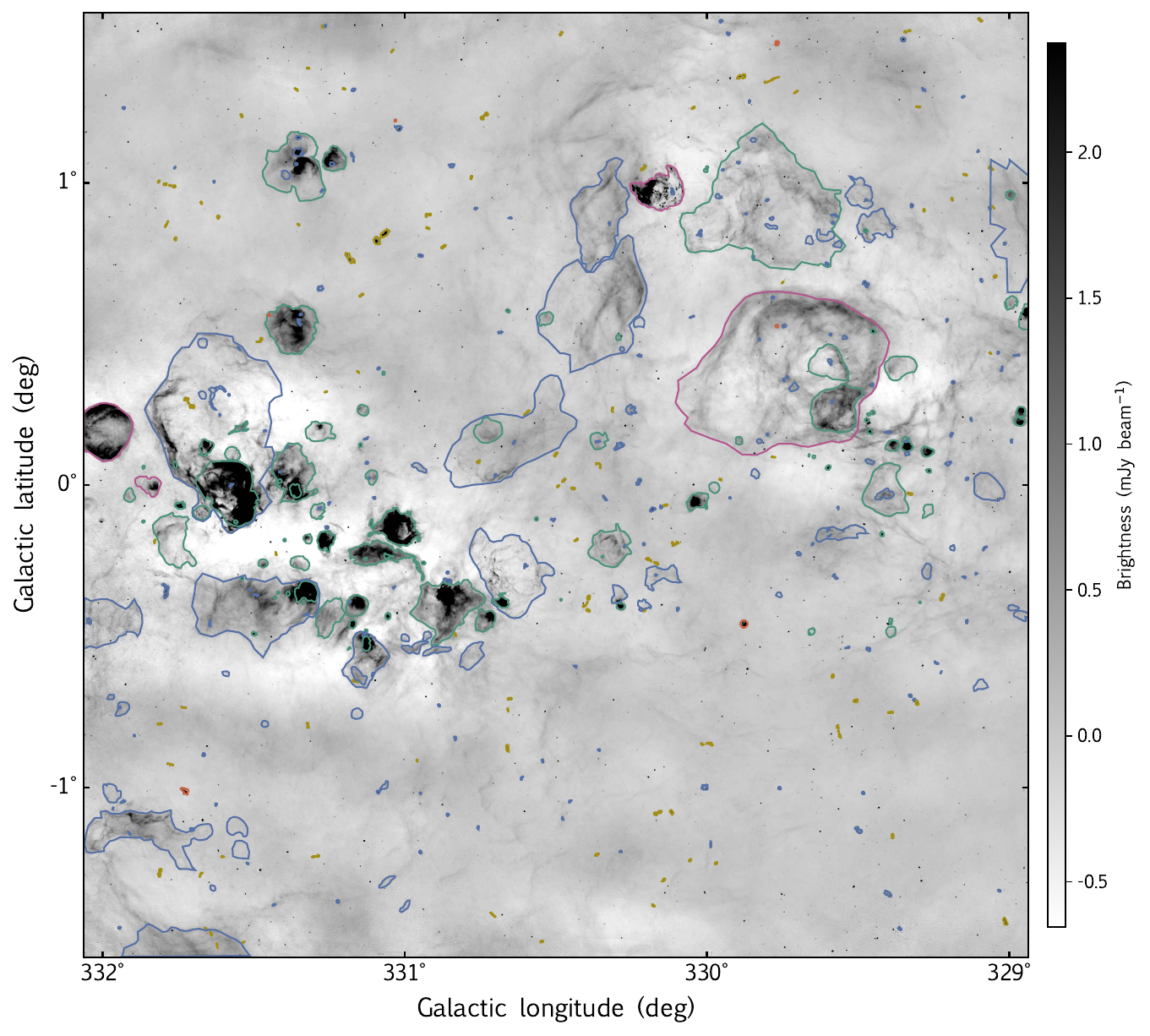}%
\caption{Example of a fully segmented tile (G330.5+0.0). The sources are coloured according to their assigned tag: magenta=SNR, bluish green=\hii{} region, orange=PN, blue=unclassified, gold=extragalactic. Isolated filamentary structures are excluded from the extended source catalogue.}
\label{fig:source-extraction}
\end{figure*}

Those DS9 regions for which no match was found in the catalogues fell into two categories:

\begin{itemize}
    \item Regions exhibiting the characteristic morphology of background active galactic nuclei (AGNs), with a bipolar, elongated shape. In some cases, these regions presented well-defined features, like core and jets. Unfortunately, most AGN catalogues in literature, such as the Unified Radio Catalogue\footnote{\url{http://www.aoc.nrao.edu/~akimball/radiocat_2.0.shtml}}\citep{2008AJ....136..684K,2014IAUS..304..238K} or the WISE AGN Catalogue \citep{2018ApJS..234...23A}, cover only extragalactic fields or deliberately avoid the Galactic plane, preventing a crossmatch with SMGPS. For this reason, we adopted a conservative approach, assigning an `EXTRAGALACTIC' tag only to the most obvious morphological candidates.
    \item Regions with morphologies inconsistent with radio galaxies, which may trace different populations of previously undetected Galactic objects. These regions were assigned the label `UNCLASSIFIED'.
\end{itemize}

\noindent A sample catalogue output produced at this stage for a representative tile (centred on Galactic longitude 330.5$^{\circ}$) is shown in Fig.~\ref{fig:source-extraction}. 

\subsection{Source characterisation}
\label{subsec:source-characterization}
A simple analysis pipeline was developed to produce a source catalogue for each tile from validated DS9 regions described in the previous section. The major steps were:

\begin{enumerate}
\item Read tile image and corresponding DS9 region data for different classes of sources (compact, extended, etc);
\item Parse DS9 region data (contour, tags, etc) and create a hierarchical tree of source objects, in which multi-islands are merged into a single object;
\item Compute source masks from input image and detected sources;
\item Compute the flux density of each extended/diffuse source and of compact sources nested inside extended/diffuse source (see Sect.~\ref{subsubsec:flux-measurement});
\item Compute other source morphological parameters;
\item Produce catalogue tables.
\end{enumerate}

\subsubsection{Flux density measurement}
\label{subsubsec:flux-measurement}

The source flux density ($F$) was computed using standard aperture photometry, scaling the background-subtracted source brightness ($S$):

\begin{equation}
    F = \frac{S}{A_\mathrm{beam}/A_\mathrm{pixel}}
,\end{equation}

\noindent where

\begin{equation}
    \frac{A_\mathrm{beam}}{A_\mathrm{pixel}} = \frac{\pi}{4\ln2}\frac{\theta_\mathrm{maj}\theta_\mathrm{min}}{(\mathrm{arcsec}^2)}\left( \frac{\mathrm{pix\ size}}{\mathrm{arcsec}}\right)^{-2}
\end{equation}

\noindent is the ratio between the area of a Gaussian beam and the pixel area. The background-subtracted source brightness $S$ is given by

\begin{equation}
    S = \sum_{i=1}^{N_{s}}S_{src,i} - \sum_{i=1}^{N_{s}}S_{bkg}
,\end{equation}

\noindent with ${N_{s}}$ the number of pixels in the source, $S_{src,i}$ the brightness of the $i$-th source pixel, and $S_{bkg}$ is the estimated background brightness. To estimate the background, we considered pixels from a rectangular \lq ring\rq\, aperture centred on the source and with a thickness of 25 pixels. We excluded from this aperture any not-a-number (NaN) pixels, as well as pixels belonging to other segmented sources (both extended and compact). Different methods were tested when estimating the background brightness:
\begin{enumerate}
\item median of background aperture, $S_{bkg,i}$=$\widehat{S}_{bkg}$, averaged over multiple concentric rectangular \lq ring\rq\, apertures;
\item $S_{bkg,i}$ set to a cubic bivariate spline interpolation\footnote{\url{https://docs.scipy.org/doc/scipy/reference/generated/scipy.interpolate.SmoothBivariateSpline.html}} for source pixel $i$, with an interpolation model obtained from the background aperture;
\item $S_{bkg,i}$ set to a 2D polynomial\footnote{\url{https://docs.astropy.org/en/stable/api/astropy.modeling.polynomial.Polynomial2D.html}} defined in $l$ and $b$ (with six terms in total) fitted for source pixel $i$, with a fit model obtained from the background aperture.
\end{enumerate}
We tested the three methods on sample image cutouts extracted from different tiles in areas free of extended and point-sources. From these, we considered different apertures to simulate artificial sources, and we estimated the background for each of them. The `true' value of the background summed up over the aperture is known in this case, as the simulated sources are made by pure background pixels. The spline and polynomial fit methods were found to provide a more accurate estimation compared to the median for a subset of the data. However, for some artificial sources, they produced rather extreme values, likely indicating a failed background modelling. For this reason, we set the default background to the more stable median estimator. We also provide the other two estimates in the catalogue table for completeness.

The statistical uncertainty on the measured flux density was computed by error propagation\footnote{See \url{https://wise2.ipac.caltech.edu/docs/release/allsky/expsup/sec2\_3f.html} for more details.}, considering the variances in the pixel fluxes within the source aperture and in the estimated background. 
Following \cite{paper1}, a systematic uncertainty of 5\% was assumed and added in quadrature to obtain the total flux density uncertainty.The reported flux density  for a given source includes possible contributions from nested sources (either point-like or extended). For completeness, the background-subtracted source flux density with compact nested sources subtracted was provided as an additional catalogue parameter (see Table~\ref{tab:catformat}).

\subsubsection{Source position and angular size}
\label{subsubsec:sourcesize}

The catalogue provides the following positional and morphological parameters for each source:

\begin{itemize}
    \item source position, estimated from source binary image moments (i.e. taking the centroid of the binary mask) and computed from the corresponding DS9 region;
    \item source signal-weighted position, computed using moments extracted from the source flux image;    \item source angular minimum and maximum extents, estimated by computing a rotated rectangle of minimum area from the source contour;
    \item source minimum bounding circle.

\end{itemize}

All values are provided both in image (pixels) and world (Galactic and equatorial) coordinates. In addition, each source is given a name following an IAU-compliant scheme (\texttt{GXXX.XXX$\pm$YY.YYY}) based on the centroid coordinates.

\subsection{Biases and limitations}
\label{subsec:biases}

Since the catalogue focuses exclusively on extended sources, it is crucial to acknowledge its limitations and potential biases to enable a proper scientific exploitation.

Firstly, the lower cut-off of 5 synthesised beam areas restricts the catalogue to sources with areas larger than $\sim$360 arcsec$^2$. In the simplest case of a perfectly circular source, such a threshold corresponds to a radius of $\sim$11 arcsec. This may introduce a significant distance-dependent selection effect across the Galaxy, with minimum source radii of $\sim$0.005--1 pc for a distance range of 0.1--20 kpc \footnote{In the range of Galactic longitude covered by the SMGPS, the Milky Way extends up to about 20 kpc from the Sun \citep{Churchwell2009}}. This effect must be carefully considered when analysing different populations, as it can be particularly critical for some source types, such as PNe, typically falling within this size range. We thus warn the reader that a complete picture of certain populations can only be achieved in combination with the complementary catalogue of compact sources (Mutale et al., in prep). The metrics presented in Sect. \ref{sec:analysis} should then be interpreted in terms of extended detections rather than absolute detection rates.

Another important limitation concerns large angular scales. Two factors influence the amount of flux density that is recovered from a given source: spatial filtering and deconvolution depth, each weighing differently depending on the source size, structure and surface brightness distribution. The theoretical LAS value of $\sim$27 arcmin corresponds to physical sizes ranging from 0.8 to 160 pc for distances between 0.1 and 20 kpc. This is the theoretical upper threshold, but the actual largest recoverable scale depends on how well the $uv$ coverage samples the spatial frequencies of the source.  However, the shallow deconvolution causes faint sources of just a few arcminutes to have incompletely recovered flux densities. Disentangling the contributions of these two effects in real scenarios is challenging. Simulations of various synthetic source profiles, provided in Appendix \ref{sec:app-LAS}, show that angular scales up to $\sim$5–10 arcminutes are generally well recovered, consistent with \cite{paper1}. An additional column has been included in the catalogue to indicate whether a source exceeds this threshold, meaning its catalogued flux density is almost certainly an underestimate.

Finally, we note that certain areas of the survey present imaging artefacts, particularly near bright sources or at the tile edges. In the most extreme cases, such artefacts can render significant portions of a given tile unusable. To ensure the quality of the catalogue, we excluded any sources located in the most problematic areas, even if they remained distinguishable. For a more detailed discussion on the instrumental and imaging issues of the SMGPS, we refer the reader to \cite{paper1}.

\section{Catalogue analysis}
\label{sec:analysis}

In this section we discuss the general distribution and statistical properties of the extended sources detected in the SMGPS. For a detailed analysis of individual objects, we refer the reader to the corresponding follow-up papers.

\subsection{General distribution and properties}
\label{subsec:distribution}

After conducting a thorough analysis of the MeerKAT tiles, we extracted a total of 16534 extended sources, of which 3293 (20\%) are tagged as diffuse.
By crossmatching the extracted sources with the ancillary catalogues described in Sect. \ref{sec:ancillary-catalogues}, a reliable identification of 24\% of the catalogued sources as Galactic objects is possible: 3323 sources matching with known \hii{} regions, 266 with SNRs, 215 with PNe, 21 with LBVs, and 7 with WR stars. In addition, 59 sources (0.4\% of the total) have ambiguous or multiple matches, presenting radio emission co-spatial and compatible with two or more catalogued sources of different nature. These numbers shall not be interpreted as an absolute detection rate: many other \hii{} regions, PNe, and evolved massive stars are detected in the SMGPS, but appear as compact sources (either point-like or with an area $< 5$ synthesised beams). These sources are not considered in the present work, but are part of the compact source catalogue (Mutale et al., in prep). The remaining extracted sources, which do not have a match in the catalogues of Galactic objects, include morphological extragalactic candidates and unclassified sources (33\% and 43\% of the total, respectively).

\subsubsection{Source density and location}

\begin{figure*}
\centering%
\includegraphics[scale=0.65]{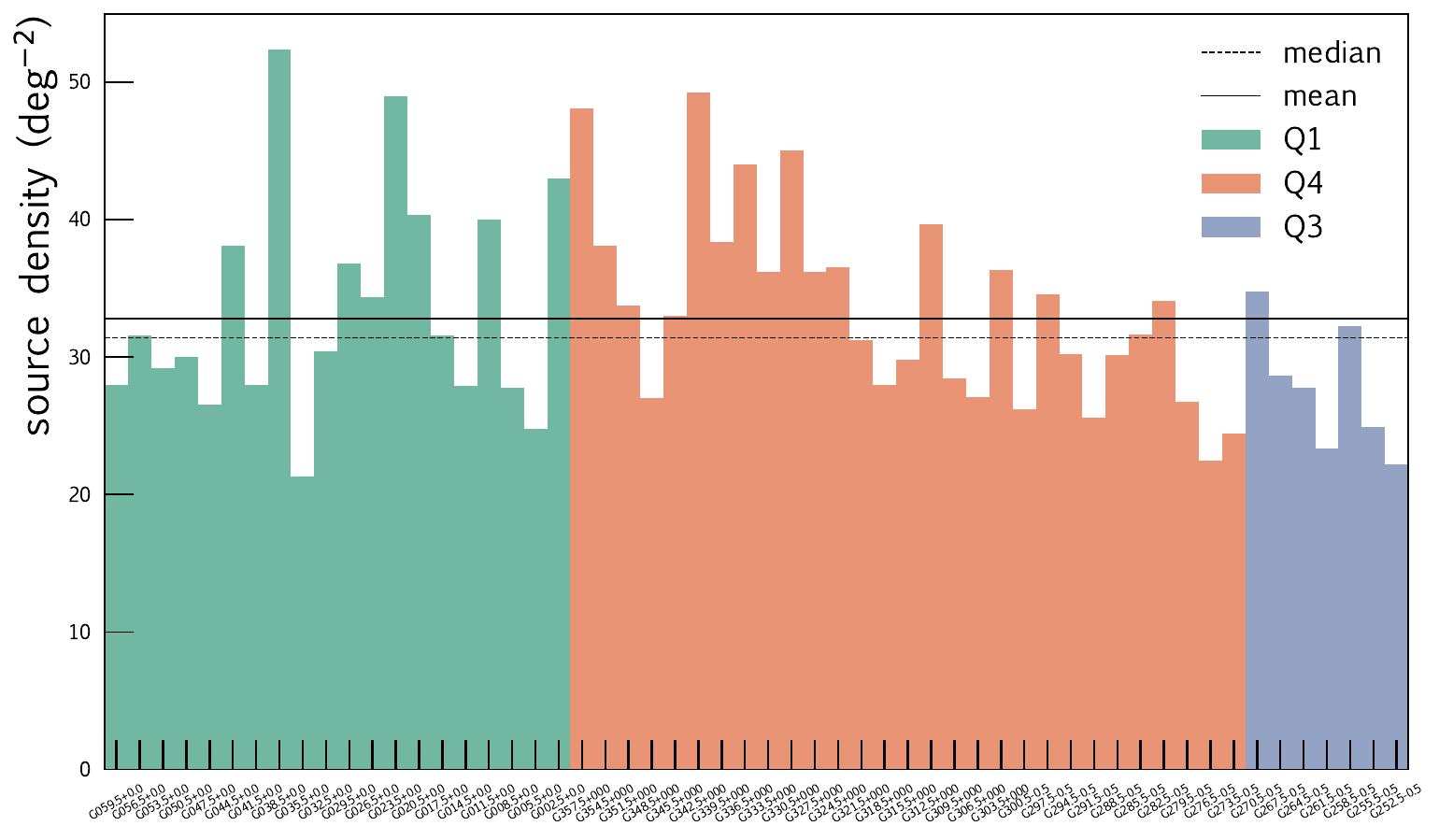}%
\caption{Extended source density per square degree for all 56 SMGPS tiles (green=quadrant 1, orange=quadrant 3, blue=quadrant 4). The solid and dashed lines represent the average and median source densities of $\sim$33 and $\sim$31 deg$^{-2}$, respectively.}
\label{fig:source-density}
\end{figure*}

The average number of extended sources per tile varies broadly across the Galactic plane, ranging from $\sim$100 to $\sim$500 (compared to $\sim$1--5$\times$10$^{4}$ compact sources per tile), with a median source density of $\sim$31 deg$^{-2}$. In Fig.~\ref{fig:source-density} we report the source density in each tile, grouped by Galactic quadrant. Density variations are on average $\sim$30\%, reaching $>$50\% for some tiles. Such variations depend on the image data quality and noise, although the main dependence seems to be tile location: a moderate increasing trend is observed towards the Galactic centre, especially in the fourth quadrant, with the median source density in the longitude range $-30^\circ < l < +30^\circ$
 deg 28\% higher than in the rest of the survey. Source counts and densities, disaggregated by source type and quadrant, are reported in Table~\ref{tab:src-count-quadrant}. 
 
\begin{figure*}
\centering
\includegraphics[scale=0.5]{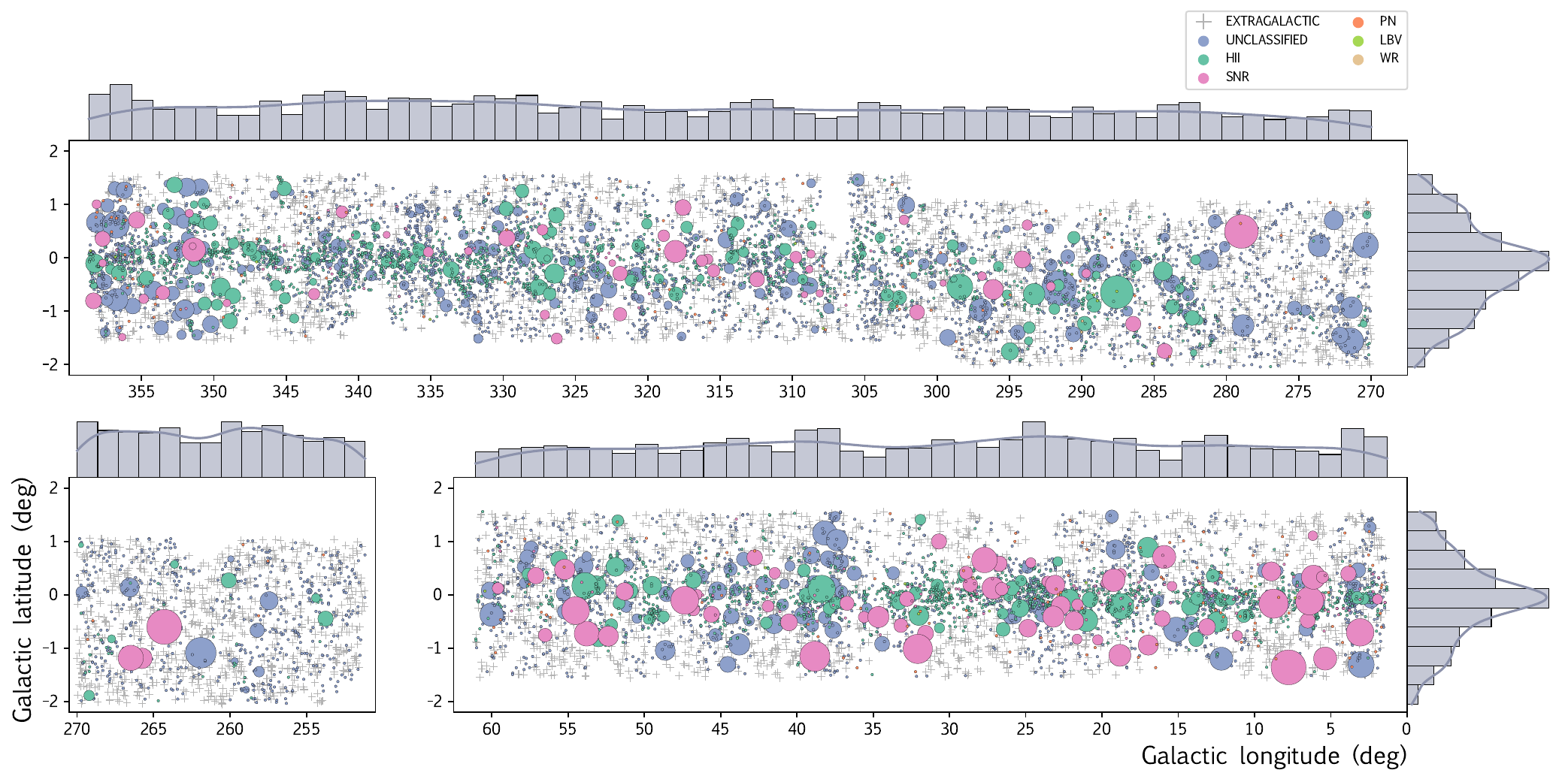}%
\caption{Sky distribution of the segmented sources in the fourth (top), third (bottom left), and first (bottom right) Galactic quadrants. Extragalactic sources are depicted as grey crosses. Galactic and unclassified sources are represented by circles, coloured by source type (blue=unclassified, bluish green=\hii{} region, magenta=SNR, orange=PN, green=LBV, and brown=WR). Sources with multiple classification tags are not shown. Note the different Galactic longitude coverage in each panel. The source distribution in the fourth quadrant shows a latitude jump around $l=300^\circ$ due to the SMGPS footprint following the Galactic warp. Sources with an angular radius of less than $10$ arcmin are represented with points of fixed size for easier visualisation. Larger sources are represented with larger circles (scaling $\propto r^{1.2}$). The histograms to the top and right of each panel represent the marginal distribution of sources in each direction (for the first and third Galactic quadrants, the cumulative latitudinal distribution is shown).}
\label{fig:sky-distribution}
\end{figure*}

 Figure~\ref{fig:sky-distribution} shows the location of all the catalogued sources in the sky, coloured by source type. Despite the different coverage in each quadrant, the crowdedness is evident in the first and fourth quadrants. Some gaps are visible in an otherwise smooth longitudinal distribution, such as the one near $l=306^{\circ}$, due to important artefacts that prevent a reliable source segmentation in significant areas.  The concentration of extended sources towards the Galactic centre also holds for some of the different source types independently, and particularly for \hii{} regions: 53\% of them are located in the ranges $l=0-30^\circ$ and $330-360^\circ$. This result is in agreement with the Galactic longitude distribution described by \citetalias{Anderson2014}.

The latitudinal source distribution in quadrants first and fourth peaks towards $b\sim0^\circ$, but the third quadrant shows a more uniform distribution. This is probably a selection effect: tiles in the third quadrant are dominated by large-scale structures (e.g. the Vela SNR, extending over three contiguous tiles; see Fig. \ref{fig:Vela}) and diffuse background emission, with a significantly reduced number of discrete \hii{} regions and SNRs. Therefore, the source distribution in these tiles is dominated by extragalactic sources, which are more easily identified  above and below the Galactic plane, far from the confusion due to foreground Galactic sources. Indeed, we observe notable statistical differences between source classes across all three quadrants: the standard deviation in $b$ is 0.47 deg for \hii{} regions and 0.52 deg for SNRs, unclassified sources show larger dispersion, with a standard deviation of 0.74 deg, and extragalactic candidates exhibit the flattest distribution, with a standard deviation of 0.93 deg.

\begin{table*}
\caption{Catalogue source numbers and relative density broken down by Galactic quadrant.}
\label{tab:src-count-quadrant}
\begin{tabular}{llllllllllll}
\hline%
\hline%
Quadrant & Area [deg$^2$]& $N_\mathrm{\hii{}}$ & $N_\mathrm{PN}$ & $N_\mathrm{SNR}$ & $N_\mathrm{WR}$ & $N_\mathrm{LBV}$  & $N_\mathrm{extragal}$ & $N_\mathrm{unclass}$ &  $N_\mathrm{multiclass}$ & $N_\mathrm{total}$ & density [deg$^{-2}$]\\
\hline%
Quadrant I      & $\sim$180     & 1534   & 105     & 170   & 2     & 8  & 1834  & 2355  & 31    & 6039  & 33.6\\%
Quadrant III    & $\sim$61      & 44     & 6       & 3     & 0     & 0  & 886   & 602   & 0     & 1541  & 25.3\\%
Quadrant IV     & $\sim$261     & 1745   & 104     & 93    & 5     & 13 & 2742  & 4224  & 28    & 8954  & 34.3\\%
\hline%
Total & $\sim$500                 & 3323   & 215     & 266    & 7   &  21 & 5462 & 7181 & 59    & 16534  & 33.1 
\end{tabular}
\tablefoot{Absolute source numbers are not directly comparable across quadrants due to their different coverage area.}
\end{table*}

\subsubsection{Source flux density}
\label{subsec:flux-validation}

\begin{figure}
\centering%
\includegraphics[width=\columnwidth]{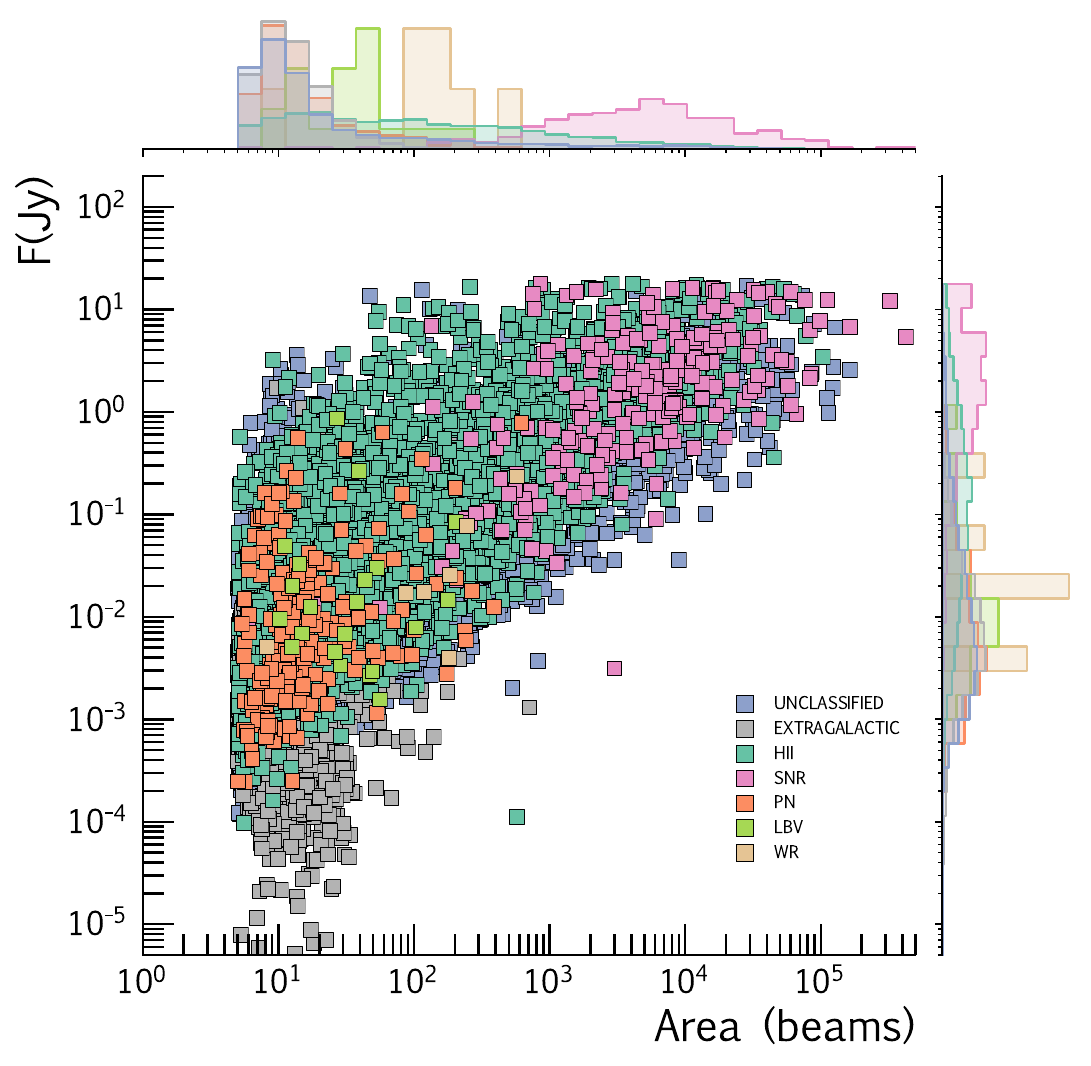}
\caption{Scatter plot of flux density as a function of the source area (in beams), coloured by source type. Histograms on the top and right show the marginal source distribution per type relative to each of the variables. Each histogram represents the normalised density function, with normalisation performed independently for each source type to facilitate the visualisation of less represented types.}
\label{fig:all_flux_vs_size}
\end{figure}

We derived flux densities for 99\% of the sources in the catalogue ($F > 0$, after subtracting the contribution from the background and child sources). As discussed in Sect. \ref{subsec:biases}, the fluxes are expected to be reliable for sources with angular sizes up to $\sim$10 arcmin, whereas the fluxes for sources larger than the LAS can be heavily underestimated. Figure \ref{fig:all_flux_vs_size} reports the flux density of the sources as a function of their area (expressed in number of beams). All source types follow a sparse distribution, with flux density increasing as the area increases. As noted in the marginal histograms, SNRs occupy the upper right region of the plot, as they are generally the largest and brightest sources, along with a fraction of \hii{} regions and very extended unclassified objects; \hii{} regions, in contrast, display a much flatter distribution, spanning several orders of magnitude in both flux and size. A similar pattern is observed with unclassified sources, which are scattered throughout the plot, but showing a trend towards smaller areas and lower flux densities, with the distributions peaking between 10 and 20 beams and 1 and 10 mJy, respectively. Extragalactic candidate sources typically correspond to the faintest, smallest sources (as expected), with flux densities as low as a few microjanskys and areas around 10 beams. Finally, PNe seem to cluster at the low-area end of the plot, with areas ranging from 10 to 20 beams, similar to structures associated with LBV or WR stars, with flux densities of the order of 50 mJy and sizes of around 1 arcmin.

\begin{figure}
\centering%
\includegraphics[width=\columnwidth]{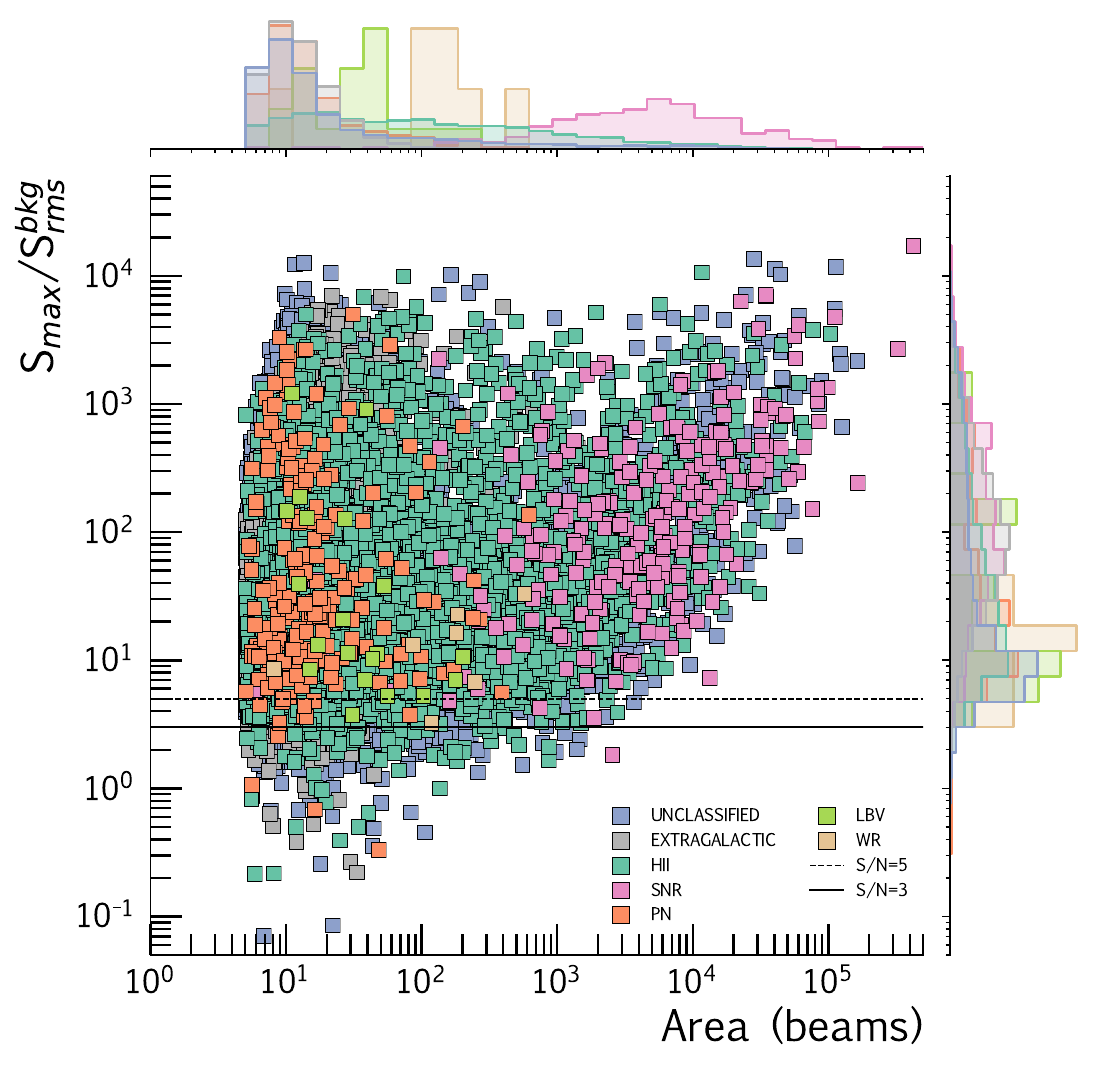}
\caption{Scatter plot of the `peak' S/N, i.e.  $S_\mathrm{max}/S_\mathrm{rms}^\mathrm{bkg}$ as a function of the source area, coloured by source type. The solid and dashed lines represent the S/N=3 and S/N=5 levels, respectively. Histograms are normalised as in Fig. \ref{fig:all_flux_vs_size}}
\label{fig:sig-noise}
\end{figure}

Since we are dealing with extended sources that may have a complex morphology and a large brightness dynamic range, providing a proper definition for the signal-to-noise ratio is tricky: the brightest filaments of large sources such as SNRs and giant \hii{} regions can be detected with a high S/N, whereas the fainter regions could be extremely shallow --- but still detectable, standing out from the diffuse background. Nonetheless, one could set an upper limit with the `peak' S/N, defined as the ratio of the peak source brightness to the local background rms (computed in concentric rectangular \lq ring\rq\, apertures surrounding the source, as described in Sect. \ref{subsubsec:flux-measurement}). The resulting distribution of peak S/N is depicted in Fig. \ref{fig:sig-noise}, as a function of the source area.

The relative source distribution in the parameter space is slightly different from that of Fig. \ref{fig:all_flux_vs_size}, with almost all source types showing a large spread in S/N. Approximately 92\% of the sources exhibit a peak S/N above 5. Conversely, only about 2\% of the sources are detected with peak brightness below the S/N=3 threshold. These sources exhibit extremely low surface brightness and typically lack well-defined boundaries, having been manually segmented  during the catalogue refinement process. Many of them belong to the unclassified category and were missed by previous, shallower surveys. Despite their faintness, they can be readily identified against the background by eye, but their derived flux parameters are likely unreliable. For the sake of completeness and to facilitate follow-up observations, we keep these sources in the catalogue.

\begin{figure}
\centering%
\includegraphics[width=\columnwidth]{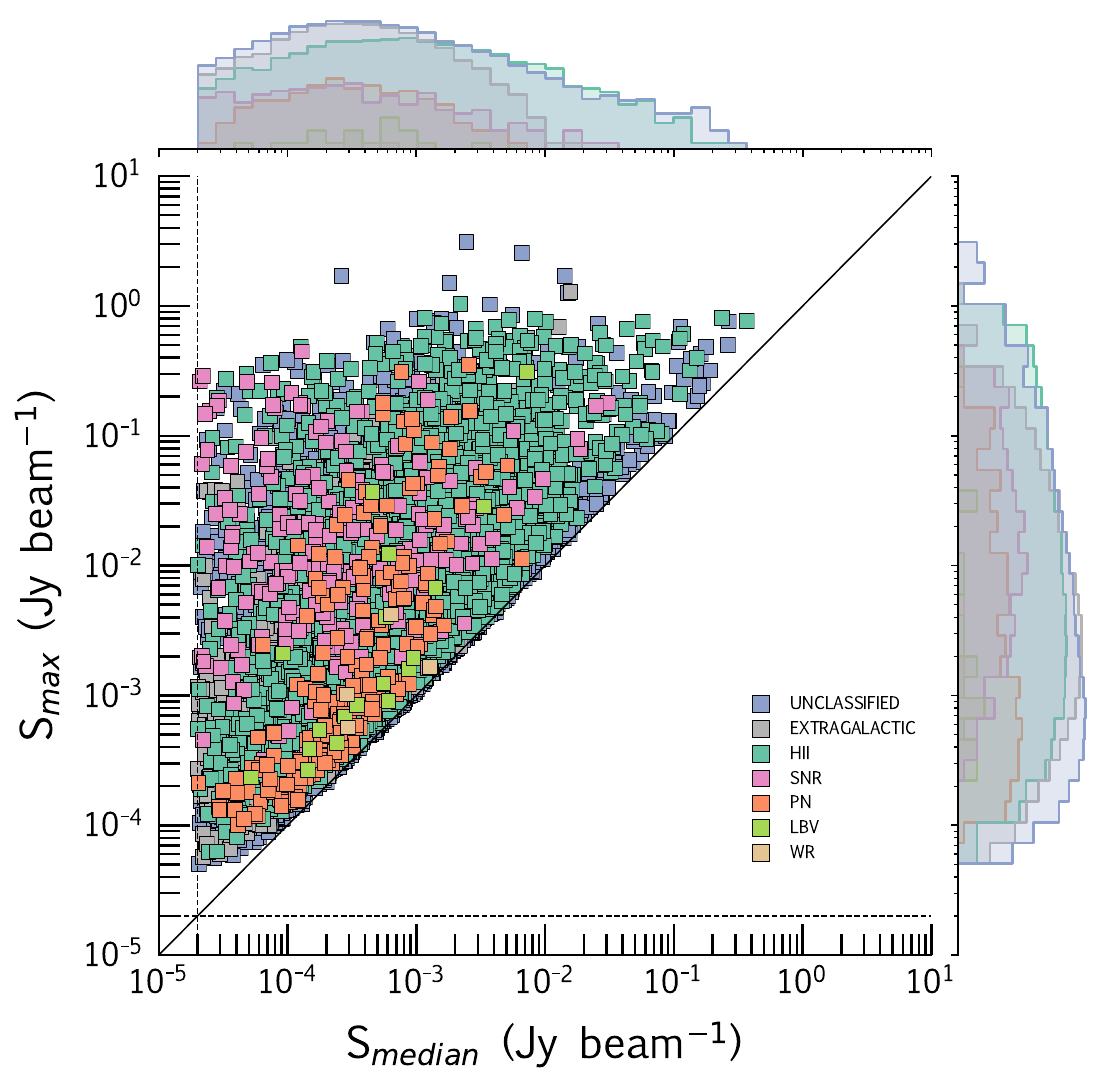}
\caption{Scatter plot of the peak and median source brightness, coloured by source type. Histograms on the top and right show the counts per source type relative to each of the variables, in log scale. The solid line represents $S_\mathrm{max} = S_\mathrm{median}$, and the dashed lines indicate the upper limit of the nominal off-source rms (10--20 $\mu$Jy beam$^{-1}$; \citealt{paper1}). Sources with median source brightness below this threshold (e.g. due to negative bowls) have been excluded from the plot.}
\label{fig:brightness-dist}
\end{figure}

It is important to remember that these values do not represent the overall S/N of the sources, but just an upper limit. Many parts, especially in the most extended sources, are significantly fainter. Figure \ref{fig:brightness-dist} shows the relative distribution of the peak source brightness, $S_\mathrm{max}$ and median source brightness, $S_\mathrm{median}$, providing insight into the uniformity of surface brightness of different source types. LBVs, WRs, and a large fraction of PNe are among the sources with the most uniform surface brightness, showing low [$S_\mathrm{max}/S_\mathrm{median}$] ratios. In contrast, SNRs exhibit the most extreme  [$S_\mathrm{max}/S_\mathrm{median}$] values, as expected given their typical radio appearance, where bright filaments stand out prominently against diffuse emission. \hii{} regions and unclassified sources, on the other hand, occupy the parameter space more evenly. Nevertheless, unclassified sources present a quite asymmetric maximum brightness distribution, peaking around $\sim$1 mJy beam$^{-1}$ and skewed towards lower values. This is again consistent with a high number of new detections within the unclassified group. The distribution of median source brightness is slightly more uniform across source types, with \hii{} regions being, on average, the brightest sources (median $S_\mathrm{median}$ $\sim$0.7 mJy beam$^{-1}$).

We assessed the reliability of the flux densities in our catalogue by comparing SMGPS flux densities with those from other surveys. Since we are dealing with extended sources, we compared with surveys that include complementary zero-baseline observations to evaluate the effect of missing short spacings. However, obtaining a fair comparison of extended sources across surveys is not straightforward, due to the intrinsic differences in resolution, sensitivity and imaging fidelity. In the first quadrant, we considered MAGPIS' VLA+Effelsberg combined image data at 1.4 GHz \citep{magpis1}, which had a significant overlap with our catalogue ($|b|<$0.8, 5$<l<$12), at a similar frequency and resolution (see Table \ref{tab:gpsurveys}).

\begin{figure}
\centering%
\includegraphics[width=\columnwidth]{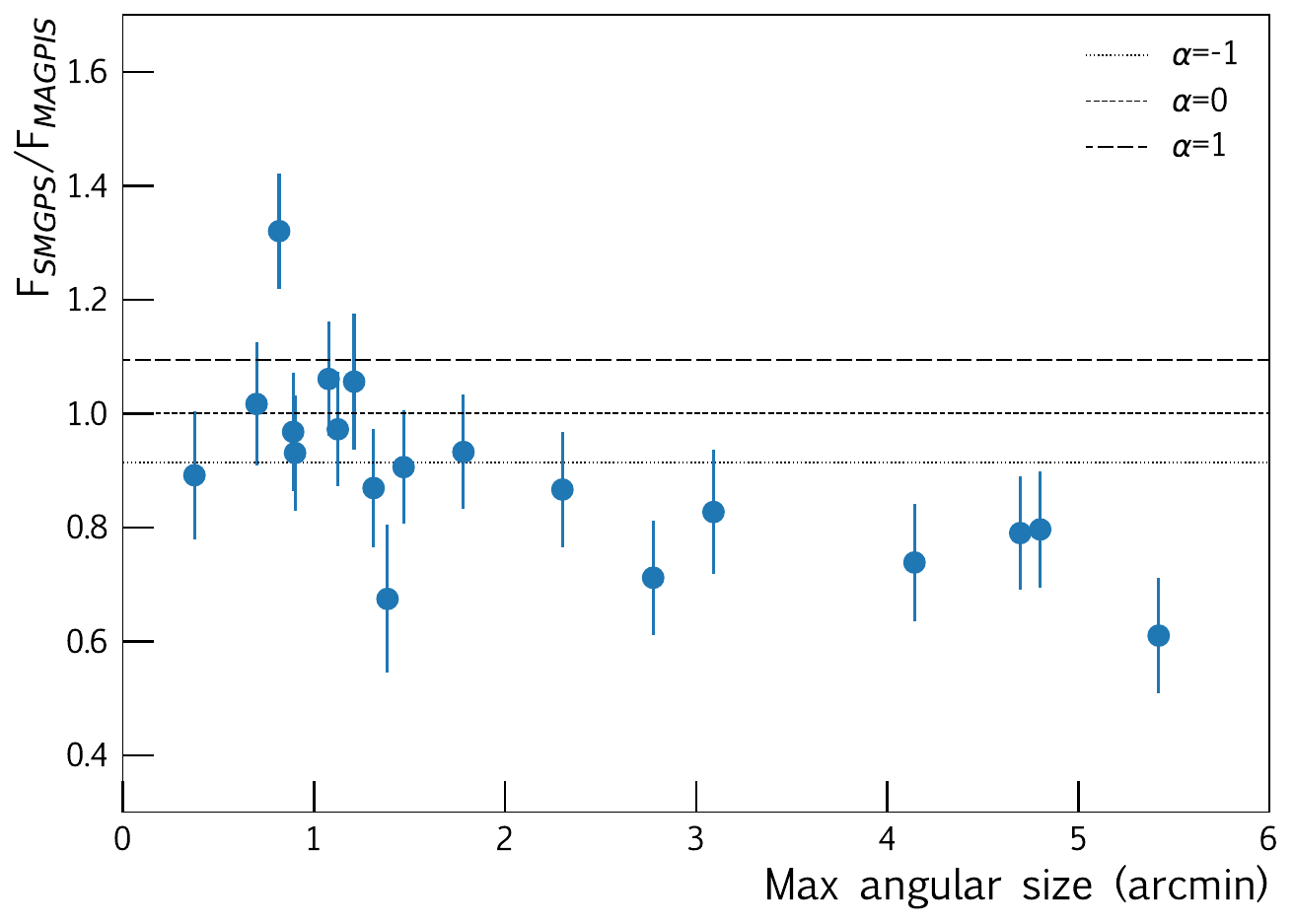}%
\caption{Comparison of the SMGPS (1.284 GHz) and MAGPIS (1.4 GHz) flux densities for selected sources (blue dots) as a function of the source maximum angular size found in MeerKAT. Dashed lines represent the expected flux ratios for sources with a power-law spectrum of spectral indices $\alpha$=$-$1,0,1.}
\label{fig:fluxratio}
\end{figure}

For the comparison, we selected a sample of isolated and bright Galactic sources to minimise contamination effects from nearby extended structures and ensure an unambiguous detection in both surveys. First, MAGPIS images\footnote{MAGPIS new 21 cm GPS data obtained from the cutout service at \url{https://third.ucllnl.org/cgi-bin/gpscutout}.} were smoothed and degraded from their original angular resolution (6.2"$\times$5.4") to the SMGPS one (8"$\times$8").
Then, MAGPIS flux densities were computed following the procedure described in Sect. \ref{subsubsec:flux-measurement} and compared with those from SMGPS. Figure~\ref{fig:fluxratio} presents the resulting [SMGPS/MAGPIS] flux density ratios for the considered sources (blue dots) as a function of their maximum angular size. Dashed lines represent the expected flux density ratios for sources with a power-law spectrum of spectral indices $\alpha$=$-$1,0,1 between SMGPS (1.284 GHz) and MAGPIS (1.4 GHz) frequencies. As shown, SMGPS flux densities can be up to $\sim$20--30\% lower than those in MAGPIS, particularly for sources larger than 2 arcminutes. This discrepancy is significantly greater than the reported SMGPS systematic uncertainty for compact sources (5\%; \citealt{paper1}) and cannot be fully attributed to the sources' spectral indices, which could only account for up to $\sim$10\% variations within the SMGPS-MAGPIS frequency range. The observed difference likely reflects a combination of spatial filtering and flux density underestimation due to the shallow deconvolution, as discussed in Sect. \ref{subsec:biases}. In addition, other factors may be at play, amplifying this effect. First, there could be residual systematics in the SMGPS data, not well understood at this early stage of MeerKAT's operational life. Second, MAGPIS flux densities may be slightly overestimated: \citealt{magpis2} reported a possible calibration issue affecting extended sources (see Fig. 6 in \citealt{magpis2} and Fig. 5 in \citealt{Kalcheva2018}). They suggested applying a background correction factor of $-$0.07 Jy arcmin$^{-2}$ to MAGPIS data to mitigate this systematic offset, likely caused by the non-trivial combination of Effelsberg and VLA observations. However, applying this correction to our MAGPIS flux densities led to excessively low values, even resulting in negative flux densities for many sources, an issue also observed in \cite{Makai2017}.

A similar flux comparison was also attempted for the fourth quadrant data with SGPS data \citep{sgps}, but unfortunately the coarse angular resolution of the SGPS survey ($\sim$2 arcmin) prevented any reliable association with the SMGPS source sample.\\

\subsubsection{Source morphology}

\begin{figure}
    \centering
    \includegraphics[width=\columnwidth]{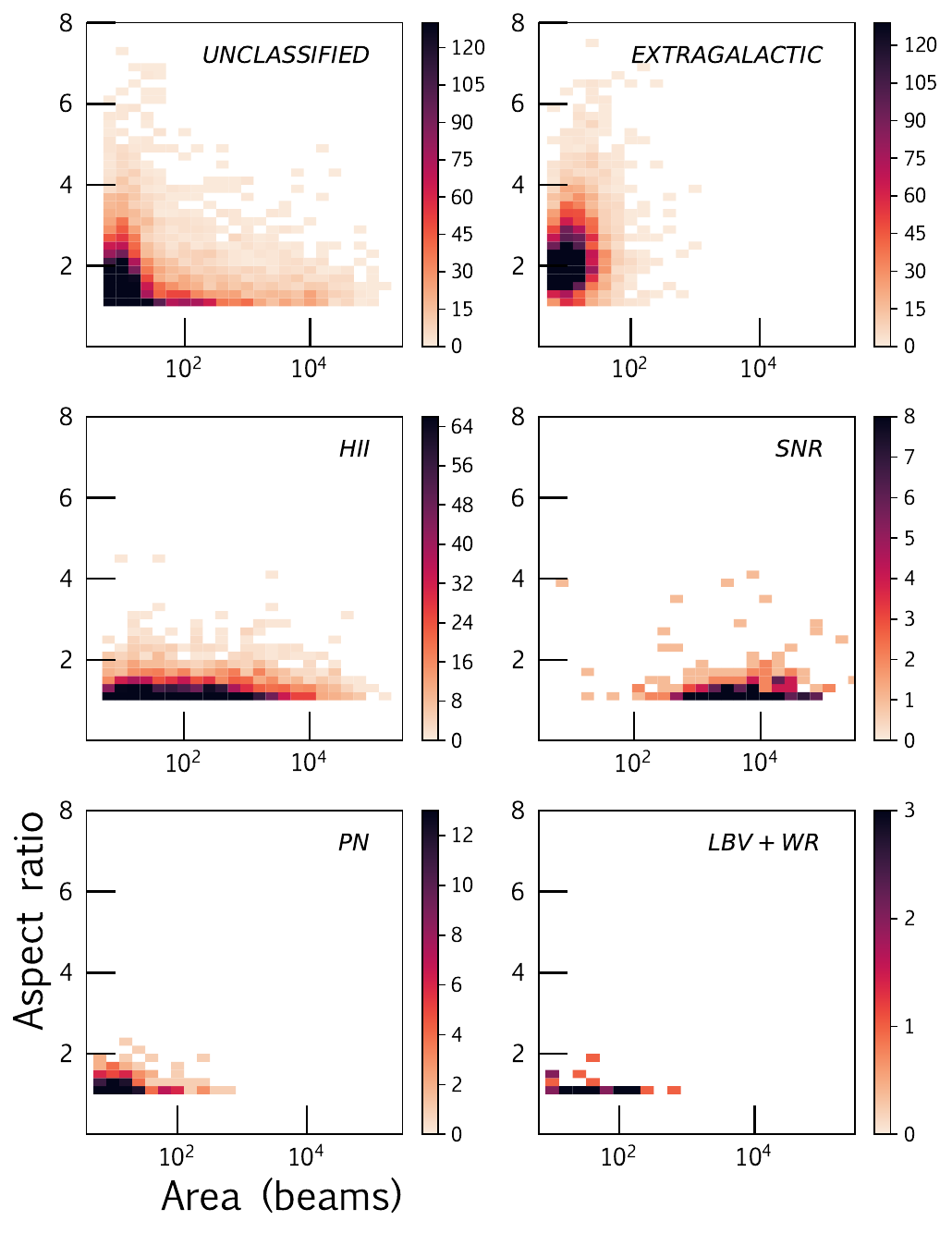}
    \caption{Bivariate histograms of the source aspect ratio (see text) versus source area. LBVs and WRs are grouped together.}
    \label{fig:aspect-ratio}
\end{figure}

Given the diverse and complex morphologies exhibited by the sources in the catalogue, a straightforward morphological analysis is not possible. Still, significant differences among source types emerge when examining their aspect ratio, defined as the ratio between the longest and shortest sides of the source rotated bounding box (see Sect. \ref{subsubsec:sourcesize}). Figure \ref{fig:aspect-ratio} shows the aspect ratio as a function of the source area for all source types. Different distributions are evident: 

\begin{itemize}
    \item SNRs and \hii{} regions tend to be roundish, showing little dispersion in aspect ratio, although they span a wide range of areas.
    \item PNe, WR and LBV are similarly roundish, with aspect ratios around $\sim$1, but they generally have much smaller areas than SNRs and most \hii{} regions. PNe show a slightly higher dispersion in aspect ratio, consistent with the bipolarity typically observed within the PNe population.
    \item Unclassified sources seem to be a superposition of different populations, with large, roundish structures as well as relatively small, elongated objects, with aspect ratios up to around 4--5.
    \item Extragalactic sources are typically smaller, and show the largest dispersion in aspect ratio, peaking around $\sim$2 and extending up to 6--7, which is compatible with their typical elongated or bipolar morphology.
\end{itemize}

\subsection{Galactic extended sources}
\label{subsec:galextended}

Approximately 24\% of the extracted extended sources can be confidently associated with known Galactic sources after a crossmatch with the source catalogues described in Sect. \ref{sec:ancillary-catalogues}.

\subsubsection{\hii{} regions}
\label{subsec:hii}

The \citetalias{Anderson2014} catalogue of \hii{} regions includes 8399 sources, with 6236 of them (74\%) located within the SMGPS coverage area. We have identified a total of 3323 extended radio sources coincident with \hii{} regions, plus 59 ambiguous matches with other object types like SNRs or PNe. Many \hii{} regions are located in densely populated areas of the Galactic plane, making it difficult to properly isolate the emission and establish an unambiguous association with a single \hii{} region in the area. For this reason, in 96 cases labelled `HII' we kept a multiple association with 225 catalogued \hii{} regions. In other words, over $\sim$3600 \hii{} regions in the \citetalias{Anderson2014} catalogue (57\% of the total in the survey area) have an extended radio counterpart in our catalogue.

\begin{figure}
\centering%
\includegraphics[width=\columnwidth]{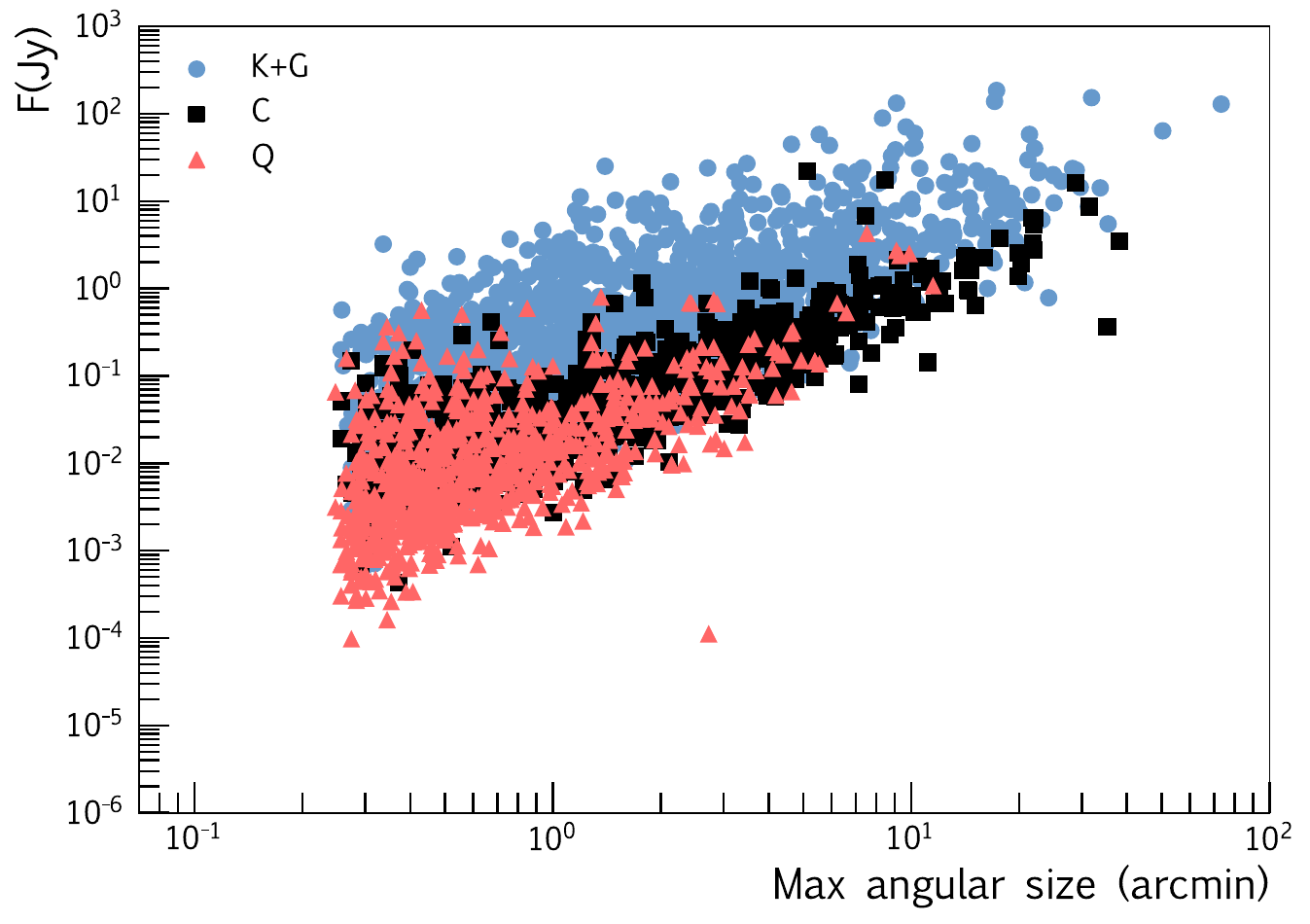}
\caption{Flux density of the detected \hii{} regions from the \citetalias{Anderson2014} catalogue, as a function of their angular size. Known sources (K) refer to those with measured radio recombination lines (RRLs) or H$\alpha$ emission. Groups (G) are H\textsc{ii} region candidates located within the same complex as a known H\textsc{ii} region. Candidate sources (C) are those co-spatial with radio continuum emission but lacking RRL or H$\alpha$ detections. Radio-quiet H\textsc{ii} regions (Q) are those for which no radio continuum detection was detected in previous surveys.}
\label{fig:hii_flux_vs_size}
\end{figure}

As for the angular size distribution, the detected \hii{} regions span two orders of magnitude, ranging from $\sim$0.2 to $\sim$75 arcmin, with an average size of $\sim$2.5 arcmin. Almost half of the sources in the \citealt{Anderson2014} catalogue are listed as \lq radio-quiet\rq, meaning they did now show detectable radio-continuum emission in previous surveys. Of these, 2940 are located within the field of the SMGPS, and we detect extended emission associated with 759 of them (25\%). Therefore, our catalogue substantially increases the number of \hii{} regions with available radio flux measurements. Figure \ref{fig:hii_flux_vs_size} shows the flux density distribution of the detected known, candidate and (previously considered) radio-quiet \hii{} regions as a function of their angular size. It can be noted that radio-quiet regions are systematically fainter and more compact than confirmed and candidates, as occurs with their infrared counterparts in \citetalias{Anderson2014}. This result is in agreement with the trend noted by \cite{Umana2021} with ASKAP Early Science data in the SCORPIO field. It suggests that for many \hii{} regions, the `radio-quietness' phenomenon is likely a matter of sensitivity rather than an intrinsic physical property, even if their apparently smaller angular size correlates with their evolutionary stage, identifying them as young, ultra-compact \hii{} regions. In fact, many more supposedly radio-quiet \hii{} regions are expected in the SMGPS compact source catalogue (Mutale et al., in prep.). In any case, the ability to detect these faint sources highlights once again the potential of MeerKAT for the study of this particular population.

\subsubsection{Evolved stars}
\label{subsec:others}

Circumstellar structures associated with evolved intermediate- and high-mass stars are rather conspicuous at infrared wavelengths, as attested by the prolific detection of infrared bubbles and shells with Spitzer \citep{Mizuno2010}, but their detection at radio wavelengths is more challenging. Indeed, the number of evolved stars with associated extended radio emission in the SMGPS is relatively low. The SMGPS covers the position of 1276 PNe, 458 WRs, and 38 LBVs, but only 215 PNe, 7 WRs, and  21 LBVs (17, 2, and 52\%, respectively) appear as extended structures with an area larger than five beams, with 18 additional sources having an ambiguous (multi-tag) classification. We note that many of the \lq missing\rq\, sources in this catalogue are instead detected as compact (smaller than five beams) and therefore belong to the compact source catalogue. Studies of these populations, including both compact and extended sources, will be presented in forthcoming science papers (Umana et al. in prep., Buemi et al., in prep., Ingallinera et al. in prep.).

The relative scarcity of extended structures larger than five beams associated with evolved stars may stem from several causes. The first and most obvious explanation is an intrinsically low surface brightness at the observing frequency. On the other hand, specially in PNe, selection effects are certainly at play, as discussed in Sect. \ref{subsec:biases}: a fraction of the PN population is known to have small angular diameters \citep{Tylenda2003}, therefore appearing as compact or unresolved radio sources in the SMGPS images, and being excluded from our catalogue. While this reasoning may also apply to some LBVs and WRs, many of these stars  display complex nebulae in optical and infrared imagery \citep{Weis2011,Toala2015}. Their ionised counterparts should typically be well resolved by the SMGPS beam, implying that the low number of extended detections, particularly for WRs, is likely due to sensitivity limitations. Furthermore, the complex background and diffuse emission near the Galactic plane complicates the detection of faint nebulae.

\subsubsection{Large-scale emission and supernova remnants}
\label{subsec:snr}
\begin{figure*}
    \centering
    \includegraphics[scale=0.9]{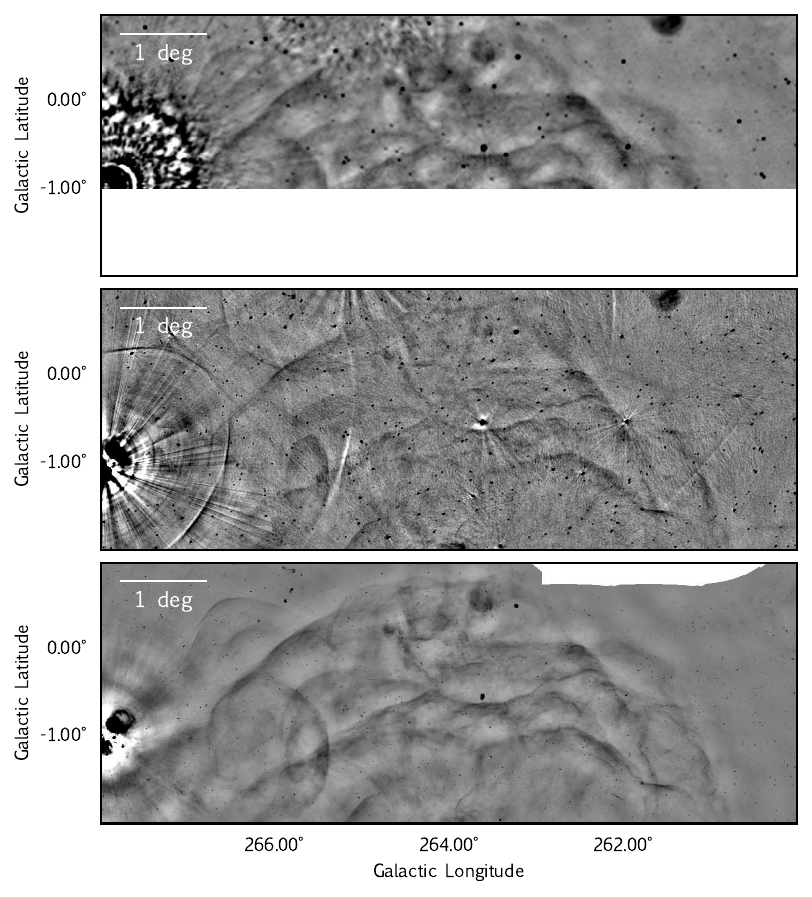}
    \caption{Vela XYZ SNR, as seen by SGPS (top), MGPS (middle), and SMGPS (bottom).}
    \label{fig:Vela}
\end{figure*}

Large-scale emission is present throughout most of the SMGPS tiles. This emission generally appears as diffuse rims and filaments, similar to those recently imaged by MeerKAT towards the Galactic centre \citep{Heywood2022}. While filamentary emission in the SMGPS is treated in detail in \cite{Williams2024}, in this work we catalogued other numerous very extended, and, in some cases, very faint structures that clearly stand out from the background. Many of these are indeed new, unclassified radio-emitters (see Sect. \ref{subsec:unknown}), while others can be linked to known Galactic SNRs due to their overall morphology and dimensions. By crossmatching the segmented sources with the SNR catalogues listed in Sect. \ref{sec:ancillary-catalogues}, we identified 266 sources unambiguously associated with SNRs, in addition to 40 others that are also co-spatial with \hii{} regions in \citetalias{Anderson2014}, for a total of $\sim$300 sources (including confirmed and candidate SNRs). For many of these SNRs, the SMGPS represents a significant leap forward in imaging quality over previously available radio data, revealing their faintest regions, and enabling a more detailed spatially resolved spectral characterisation. Indeed, a study of the radio morphology of known and candidate SNRs in the SMGPS is presented in \cite{Anderson25}, and a spectral analysis of a subsample of 29 SNRs is reported in \cite{Loru24}. 

The Vela SNR, spreading across three adjacent tiles, is also detected, as shown in Fig. \ref{fig:Vela}. This source is an excellent example to illustrate MeerKAT's ability to homogeneously image extremely extended structures while retaining fine spatial details. Comparison with images of the same area from MGPS and SGPS, clearly demonstrates the improvement in angular resolution and sensitivity, despite the impact of missing short spacings on the total recovered flux.

\subsection{Candidate extragalactic sources}

\begin{figure*}
\includegraphics[width=\textwidth]{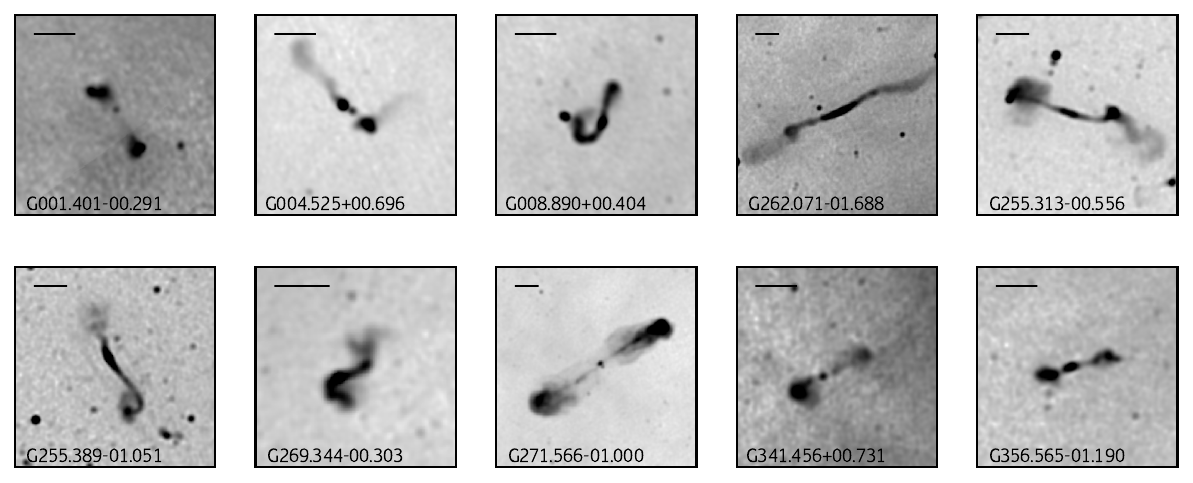}%
\caption{Examples of SMGPS sources labelled `EXTRAGALACTIC' due to their morphological features. The catalogue source name appears in the bottom-left corner of each panel, and the scale bar in the top-left corner indicates 1 arcmin.}
\label{fig:radio-galaxies}
\end{figure*}

Around 33\% of the sources in the catalogue are classified as candidate radio galaxies based on their characteristic morphology, as discussed in Sect. \ref{subsec:source-crossmatch}. These sources display a bipolar or peanut-shaped structure and, in some cases, clearly resolved components. The largest --- and thus best resolved --- sources exhibit more intricate patterns, including curved or swirling jets and strongly asymmetric brightness distributions typically seen in AGNs. These features explain the aspect ratio distribution shown in Fig. \ref{fig:aspect-ratio}. Some examples of these extragalactic sources are displayed in Fig.~\ref{fig:radio-galaxies}.

We attempted to compare the density of candidate extragalactic sources in the SMGPS, of $\sim$10.9 deg$^{-2}$, with that observed in extragalactic fields. For instance, the Early Science catalogue of the MIGHTEE survey \citep{Hey22b}, which is an order of magnitude deeper than SMGPS, reports 898 and 1376 \lq resolved\rq\, sources in the COSMOS (1.6 deg$^{2}$) and XMM-LSS (3.5 deg$^{2}$) fields, corresponding to an average source density of $\sim$450 deg$^{-2}$. If we instead limit the MIGHTEE catalogue to $3\sigma$ detections at the SMGPS sensitivity, the source density drops to $\sim$230 deg$^{-2}$. However, these figures do not make a fair comparison because our selection criteria would lead to excluding numerous resolved yet compact MIGHTEE sources. On the other hand, \cite{Gupta24} extracted 2800 extended radio galaxies from ten tiles of ASKAP's EMU Pilot Survey, covering an area of ~270 deg$^2$ (observed at 944 MHz, 16 arcsec resolution). The resulting source density, of $\sim$10.4 deg$^{-2}$, is close to the SMGPS value, but again the difference in synthesised beam needs to be considered: many \lq compact\rq\, radio galaxies in the EMU-PS field would meet the inclusion criteria of our catalogue if observed with a synthesised beam of 8 arcsec.

In any case, considering the scarcity of extragalactic surveys covering the extent of the SMGPS, it is likely that many of the candidate extragalactic sources proposed in the catalogue are new. We conducted a search within the NASA/IPAC Extragalactic Database\footnote{\url{https://ned.ipac.caltech.edu/}} (NED), using a search radius of 5 arcsec around the source centroids. We found that approximately 50\% of the sources have at least one match in the NED database. Among these, only a $0.6\%$ have a matching entry identified as a galaxy (i.e. classified as `\texttt{G}', `\texttt{QSO}', or related types), while the remaining matches are mostly either infrared or radio sources (type `\texttt{IrS}' or `\texttt{RadioS}', respectively), without any confirmation of their nature. This underscores the possibilities of this catalogue for extragalactic science.

\begin{figure*}
\includegraphics[width=\textwidth]{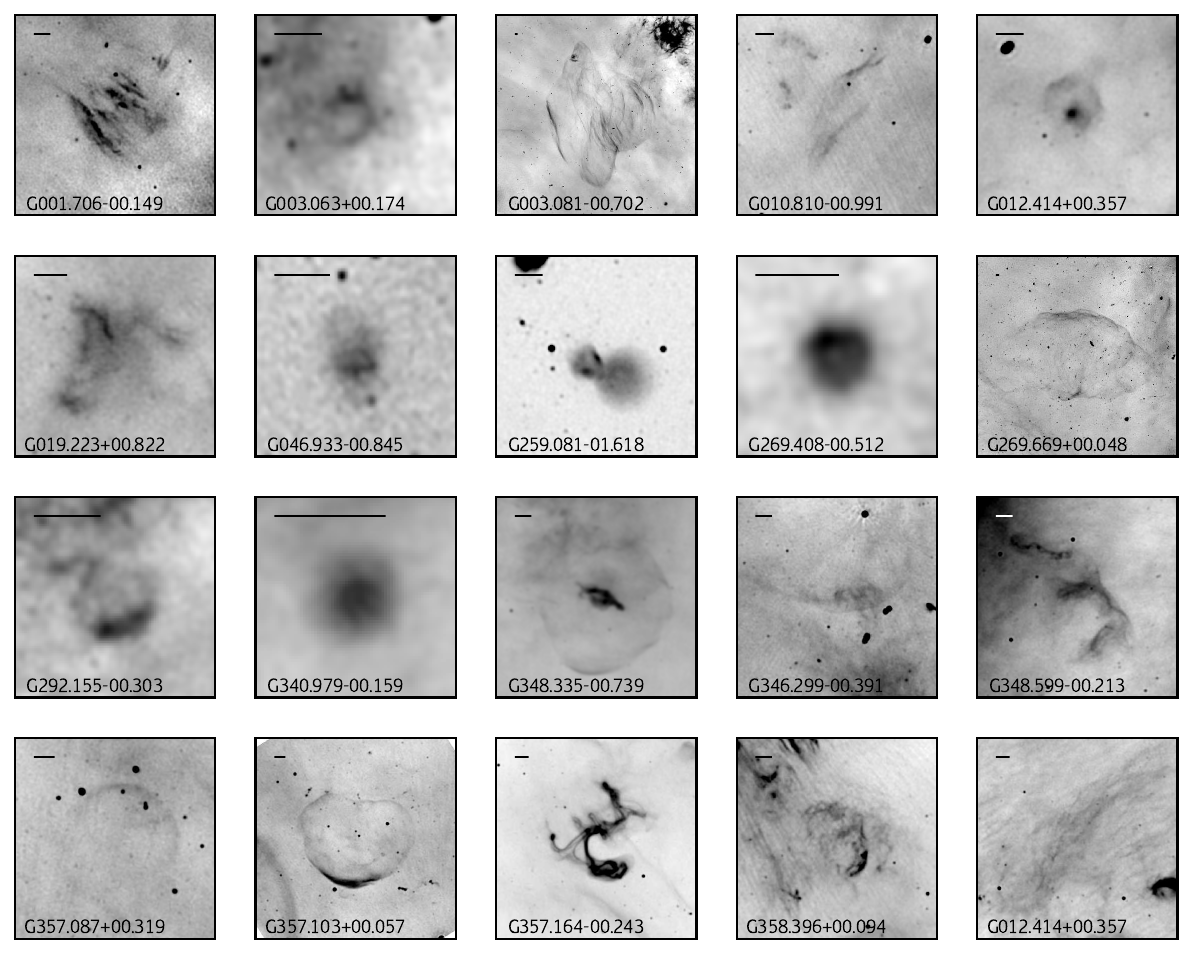}%
\caption{Examples of SMGPS sources labelled `UNCLASSIFIED'. The catalogue source name appears in the bottom-left corner of each panel, and the scale bar in the top-left corner indicates 1 arcmin.}
\label{fig:unk-examples}
\end{figure*}

\subsection{Unclassified sources}
\label{subsec:unknown}

About 43\% of the sources neither have a match in the ancillary catalogues of Galactic objects listed in Sect. \ref{sec:ancillary-catalogues} nor display the characteristic features of an AGN. Therefore, these sources have been labelled `unclassified'.

Unclassified sources constitute a heterogeneous group, with a wide variety of angular sizes and morphologies. Considering the conservative approach taken in identifying extragalactic candidates, detailed in Sect. \ref{subsec:source-crossmatch}, it is possible that a subset of unclassified sources still have an extragalactic nature. However, many of these sources could be new Galactic objects. A few representative examples of unclassified sources are displayed in Fig. \ref{fig:unk-examples}.  Looking at their range of morphologies, they can be naively divided into four groups:
\begin{itemize}
\item Complex diffuse structures, possibly part of larger structures that either have been resolved out by the interferometer or cannot be disentangled from the background due to confusion. Some examples are \texttt{G010.810-00.991}, \texttt{G012.414+00.357}, and \texttt{G019.223+00.822}.
\item Irregular sources with intricate morphologies, which may be part of larger structures that cannot be properly segmented (e.g. bright arcs of \hii{} regions in crowded tiles), or even more exotic objects, such as \texttt{G357.164-00.243} (dubbed the `heartworm nebula'; \citealt{Cotton22}) and \texttt{G358.396+00.094}.
\item Large-scale, diffuse bubble-like sources or shells, most likely new giant \hii{} regions or SNRs, like \texttt{G003.081-00.702}, \texttt{G269.669+00.048}, \texttt{G357.087+00.319}, \texttt{G357.103+00.057}, or \texttt{G348.335-00.739}. Using mid-infrared data from GLIMPSE, Spitzer and WISE, \cite{Anderson25} have identified 237 unclassified sources in the SMGPS as new SNR candidates.
\item More compact bubbles and shells, of $\sim$1 arcmin or less, possibly new \hii{} regions or nebular structures, relics of the mass-loss events of evolved stars. \texttt{G012.414+00.357}, \texttt{G269.408-00.512}, \texttt{G292.155-00.303}, and \texttt{G340.979-00.159} fall within this category.
\end{itemize}

These cases are not meant to be exhaustive, and the proposed classifications have not been confirmed. Many other interesting objects will likely be discovered upon thorough scrutiny of the full catalogue. Providing a tentative classification for these unclassified sources is far beyond the scope of this paper, and will require complementary multi-wavelength data, spectral index analysis, and probably follow-up observations. If anything, the aforementioned examples illustrate how MeerKAT can make substantial contributions to the census of Galactic radio emitters, while also leading to new, unexpected discoveries.

\section{Conclusions and future work}
\label{sec:summary}

In this paper we have presented a catalogue of extended radio sources from the SMGPS. Starting from the original SMGPS data products, we performed a semi-automated segmentation to extract extended and diffuse radio sources with areas larger than five synthesised beams. Then, the extracted sources were crossmatched with catalogues of known Galactic objects and labelled accordingly. Finally, to enrich the catalogue, the sources were ingested into a simple analysis pipeline to derive positional, morphological, and flux-related parameters. 

The final catalogue comprises a total of 16534  extended radio sources, 24\%  of which  are confidently linked to known Galactic objects (3323 sources associated with \hii{} regions, 266 with SNRs, 215 with PNe, 21 with LBVs, 7 with WRs, and 59 with multiple associations). The rest correspond to candidate extragalactic sources (33\%) or unclassified objects (43\%). 

The analysis of the catalogue yielded interesting insights into the statistical properties of the studied populations, in particular notable differences in size, surface brightness, and flux density between source types. Remarkably, unclassified sources constitute the largest group in the catalogue. A significant fraction of them are very faint and have not been detected in previous radio surveys. In this respect, MeerKAT's superb sensitivity will certainly lead to many serendipitous findings, like odd radio circles \citep{Norris2021} and similar structures \citep{Bordiu24}.

The production of this catalogue has also highlighted the inherent issues and limitations of generating comprehensive source catalogues in the SKA precursor era, namely: the inability of automated source finders to deal with faint extended sources and irregular morphologies; the need for a time-consuming visual inspection and manual refinement; and the risks and potential biases of human-based source segmentation and crossmatch identification (for instance, poor segmentation and spurious associations in heavily crowded areas).

 All things considered, MeerKAT is fuelling a revolution in Galactic science with its exceptional imaging capabilities, and the SMGPS represents a window to the future of Galactic plane surveys. The extended source catalogue presented in this work, arguably one of the largest and deepest to date, will be a valuable resource for those interested in studying radio continuum sources in the Milky Way. Similarly, the extragalactic community will benefit from the extensive number of catalogued AGN candidates at low Galactic latitudes --- a region barely explored by extragalactic radio surveys.

\section*{Data availability}
\label{sec:app-data-availability}
The full SMGPS extended source catalogue is hosted at \url{https://doi.org/10.48479/t1ya-na33} and is also available at the CDS via anonymous ftp to \url{cdsarc.u-strasbg.fr} (130.79.128.5) or via \url{http://cdsweb.u-strasbg.fr/cgi-bin/qcat?J/A+A/}. All SMGPS DR1 data products are available at \url{https://doi.org/10.48479/3wfd-e270}.

\begin{acknowledgements}
We thank the anonymous referee for their detailed and valuable comments that improved the quality of the paper. The MeerKAT telescope is operated by the South African Radio
Astronomy Observatory, which is a facility of the National Research
Foundation, an agency of the Department of Science and Innovation.
The National Radio Astronomy Observatory is a facility of the National Science Foundation operated under cooperative agreement by
Associated Universities, Inc. The Centre for Astrophysics Research at
the University of Hertfordshire kindly provided access to their HPC
facilities for data processing and storage.
This work was supported in part by the Italian Ministry of Foreign Affairs and International Cooperation, grant number ZA23GR03. We thank the authors of the following software tools and libraries that have been extensively used for data reduction, analysis, and visualization: \caesar{} \citep{Riggi2016,Riggi2019}, astropy \citep{astropy2013,astropy2018}, \textsc{Root} \citep{Brun1996}, \textsc{Topcat} \citep{Taylor2005, Taylor2011}, ds9 \citep{Joye2003}, APLpy \citep{Robitaille2012}.\\
S.R. benefited from the "CIRASA" and "SCIARADA" INAF research grants. 
C.B. benefited from grant No. 863448 (NEANIAS) of the Horizon 2020 European Commission programme.
M.A.T gratefully acknowledges the support of the Science \& Technology Facilities Council through grant awards ST/R000905/1 and ST/W00125X/1. 
\end{acknowledgements}

%
%

\bibliographystyle{aa}
\bibliography{references}

\begin{appendix}

\section{Flux density recovery}
\label{sec:app-LAS}

\begin{figure}
\centering%
\includegraphics[width=0.95\columnwidth]{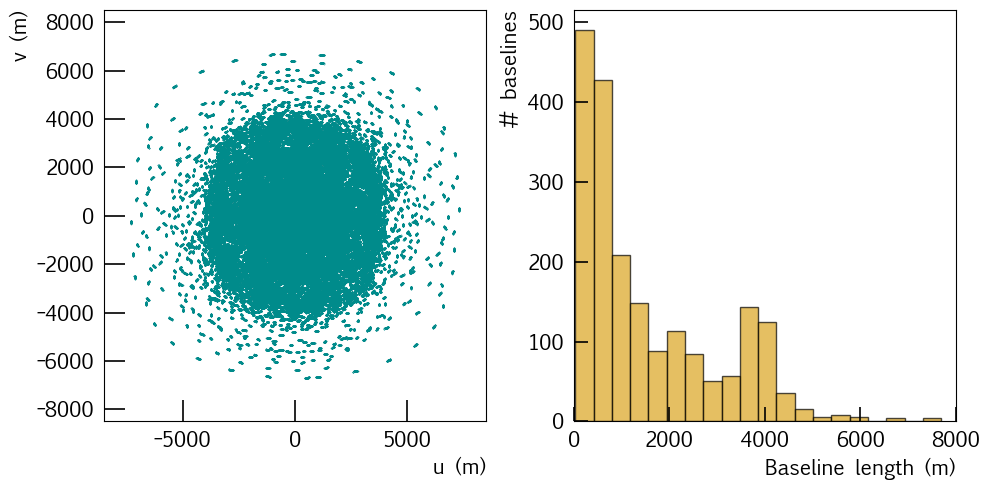}%
\caption{Top: Simulated MeerKAT $uv$ coverage for a $\sim$1h pointing at $\delta=-30$ deg, split over 10 hours to approximate the real SMGPS coverage. Bottom: Histogram of baseline lengths ($N$=2016 baselines).}
\label{fig:uv-coverage}
\end{figure}

To investigate the impact of spatial filtering and shallow deconvolution on the accuracy of flux recovery for extended sources, we simulated MeerKAT observations of various source profiles, covering a range of angular scales and flux densities.  The simulations were carried out using CASA\footnote{https://casa.nrao.edu/} task \texttt{simobserve}, with antenna positions and primary beam information retrieved from the MeerKAT wiki\footnote{\url{https://skaafrica.atlassian.net/servicedesk/customer/portal/1/article/277315585}}. Thermal noise was not included. To approximate the actual $uv$ coverage of the SMGPS observations  we followed a similar observing strategy \citep{paper1}, performing a 7.5 min scan every hour for 10 hours (see Fig. \ref{fig:uv-coverage})
All the sources were given a fixed declination $\delta=-30$\degr.

We imaged the visibilities resulting from each simulated observation using the CASA task \texttt{tclean}. To ensure a fair comparison with the SMGPS data, we applied a shallow deconvolution using only point \textsc{clean} components to a maximum depth of 100 $\mu$Jy beam$^{-1}$. Finally, the flux density of the sources was measured in the image plane using aperture photometry.

\begin{figure}
\centering%
\includegraphics[width=0.95\columnwidth]{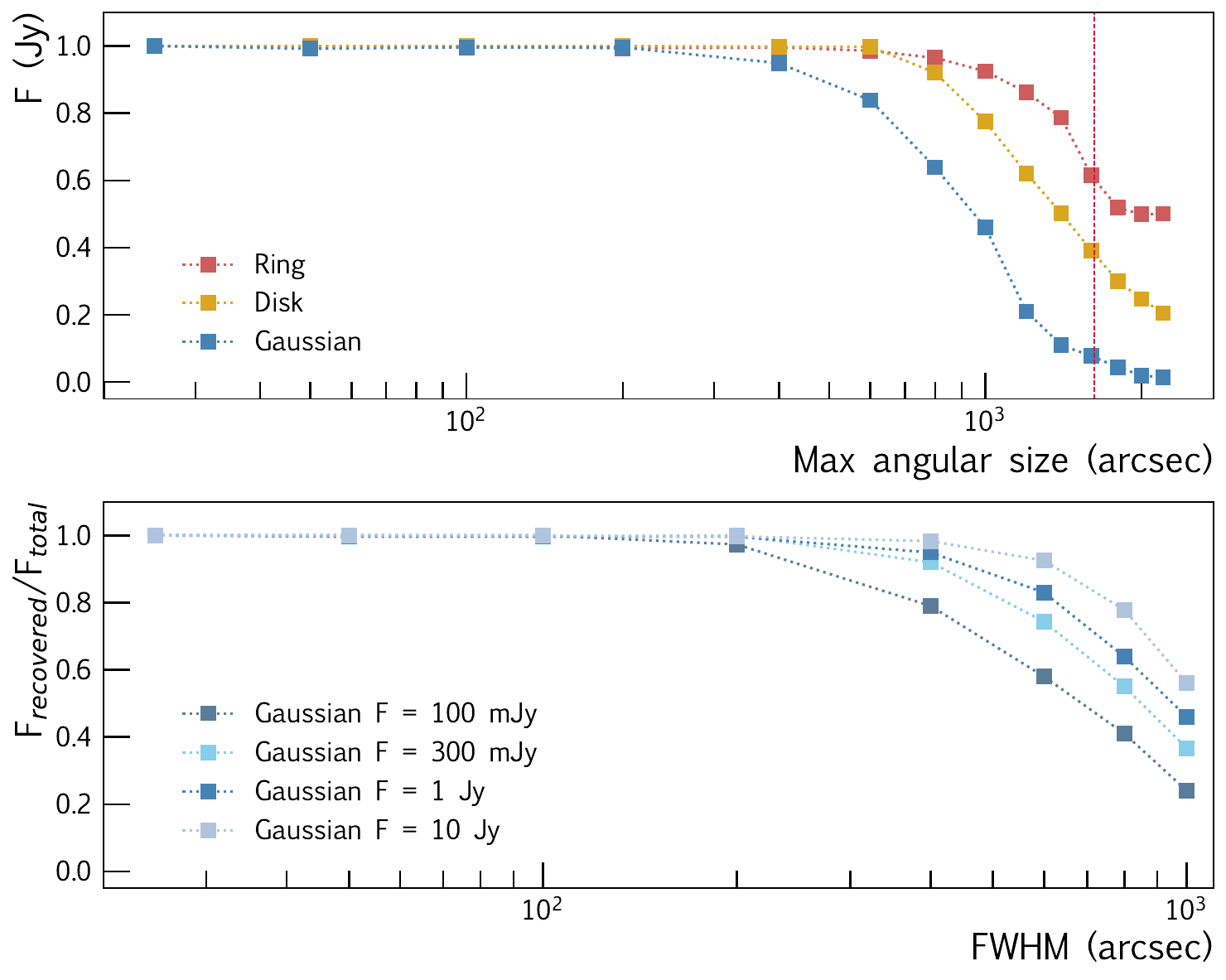}%
\caption{Top: Recovered flux density as a function of the maximum angular size for a uniform disk, a ring, and a Gaussian source of $F=1$ Jy. The dashed vertical line represents 27 arcmin, the theoretical LAS at the representative frequency of the maps. Maximum angular size refers to the diameter (disk), outer diameter (ring), and full width at half maximum (Gaussian). Bottom: Flux density recovery performance for Gaussian sources of 100 mJy, 300 mJy, 1 Jy, and 10 Jy.}
\label{fig:fig-LAS}
\end{figure}

Figure \ref{fig:fig-LAS}, top panel, shows the recovered flux density as a function of angular size for three types of sources: uniform disks, rings (with a thickness of 20\% of the radius), and Gaussians. The sources span angular scales up to 2200 arcseconds ($\sim$37 arcminutes) and have a fixed flux density of 1 Jy (meaning the flux is spread over a larger surface area as the source size increases). In all cases, we see that the recovered flux density drops significantly before reaching the theoretical LAS (indicated by the dashed red line).

The fraction of recovered flux density is strongly influenced by the source structure, specifically how the flux is distributed across spatial scales. For a perfect ring, the flux density is concentrated on the highest spatial frequencies (the \lq edges\rq), so the flux density is reasonably well recovered even for large sources. In contrast, in a uniform disk the flux is evenly distributed across the entire emission area, with the \lq flat\rq\, portion more easily resolved out and leading to a faster drop-off. Finally, Gaussian sources, having a smoother surface brightness distribution, represent the least favourable scenario as their structure becomes dominated by lower spatial frequencies, showing substantial flux density loss beyond scales of $\sim$10 arcminutes.

The sources in these simulations are bright (1 Jy), and in this regime flux recovery is inherently better due to the reduced impact of shallow deconvolution. For fainter sources, however, shallow deconvolution becomes increasingly problematic. To explore this further, we performed a second set of simulations using Gaussian sources with integrated flux densities of 100 mJy, 300 mJy, 1 Jy, and 10 Jy and increasing full width at half maximum. The results, in Fig. \ref{fig:fig-LAS}, bottom panel, show that the scale at which flux density is reliably recovered depends critically on the source brightness (specifically, the fraction of the source whose surface brightness falls below the deconvolution threshold). For a source of 100 mJy, flux loss may exceed $\sim$20$\%$ at scales as small as 4-5 arcmin.

These simulations indicate that shallow deconvolution is the primary obstacle to flux density recovery in the low-surface brightness regime. This effect, which could worsen in real sources with more complex structures, likely contributes to the discrepancies observed in Fig. \ref{fig:fluxratio}. In summary, considering the contributions of spatial filtering and shallow deconvolution, SMGPS flux densities at angular scales larger than $\sim$5–10 arcminutes should be interpreted with great caution, in agreement with the recommendations of \cite{paper1}.

\label{app:catalogue}
\onecolumn

\section{Catalogue description}
\label{sec:app-cat-description}

\begin{table*}[ht!]
\caption{Format of the SMGPS extended source catalogue (64 data columns per source).}              
\label{tab:catformat}      
\footnotesize
\centering                                      
\begin{tabular}{c | l | c | l}          
\hline%
\hline%
Col. Num. & Name & Unit & Description\\%
\hline%
1 & \emph{tile\_name} & - & Tile image name\\%
2 & \emph{sname} & - & Source name\\%
3 & \emph{sname\_iau} & - & Source name in IAU format\\%
4 & \emph{npix} & - & Number of pixels in island\\%
5 & \emph{nnested} & - & Number of child (nested) sources\\%
6 & \emph{nnested\_ext} & - & Number of extended nested sources \\%
7 & \emph{nested\_ext} & - & Source name of the extended nested sources \\%
8 & \emph{morph\_type} & - & Morphology tag \{0=\texttt{EXTENDED}, 1=\texttt{DIFFUSE}\}\\%
9-10 & (\emph{bbox\_x},\emph{bbox\_y}) & - & Source bounding box centre in pixels\\%
11-12 & (\emph{bbox\_w},\emph{bbox\_h}) & - & Source bounding box width and height in pixels\\%
13-14 & (\emph{bbox\_rot\_x},\emph{bbox\_rot\_y}) & -,- & Source rotated bounding box centre in pixels\\%
15-16 & (\emph{bbox\_rot\_w},\emph{bbox\_rot\_h}) & -,- & Source rotated bounding box width, height in pixels and rotation angle\\%
17 & \emph{bbox\_theta} & deg & Source rotated bounding box angle\\%
18-19 & (\emph{x},\emph{y}) & - & Source centroid position in pixels\\%
20-21 & (\emph{l},\emph{b}) & deg,deg & Source centroid position in Galactic coordinates\\%
22-23 & (\emph{$\alpha$},\emph{$\delta$}) & deg,deg & Source centroid position in J2000 Equatorial coordinates \\%
24-25 & (\emph{x\_w},\emph{y\_w}) & - & Source flux-weighted centroid position in pixels\\%
26-27 & (\emph{l\_w},\emph{b\_w}) & deg,deg & Source flux-weighted centroid position in Galactic coordinates\\%
28 & \emph{radius} & - & Half the diagonal in pixels of the source bounding box\\%
29 & \emph{radius\_wcs} & arcmin & Half the diagonal of the source bounding box\\%
30-31 & (\emph{x\_minCircle},\emph{y\_minCircle}) & - & Minimum circle centroid position in pixels\\%
32 & \emph{radius\_minCircle} & - & Radius in pixels of source minimum bounding circle\\
33-34 & (\emph{x\_minCircle\_wcs},\emph{y\_minCircle\_wcs}) & deg,deg & Minimum circle centroid position in Galactic coordinates \\
35  & \emph{radius\_minCircle\_wcs} & arcmin & Radius of source minimum bounding circle \\ 
36-37 & (\emph{minSize\_wcs,maxSize\_wcs}) & arcmin & Source angular minimum and maximum size obtained from minimum\\
& & &  bounding rectangle\\%
38 & \emph{S} & Jy/beam & Sum of source pixel brightness\\%
39-40 & (\emph{S$_{min}$,S$_{max}$}) & Jy/beam & Minimum and maximum pixel brightness in source\\%
41-43 & (\emph{S$_{mean}$},\emph{S$_{median}$},\emph{S$_{rms}$}) & Jy/beam & Mean, median and standard deviation of source pixel brightness\\%
44-46 & (\emph{S$_{bkg}$},\emph{S$_{bkg}^{interp}$},\emph{S$_{bkg}^{fit}$}) & Jy/beam & Sum of background brightness extrapolated over source pixels from median, \\
& & & interpolation, and polynomial fit methods\\%
47 & \emph{S$_{err}^{bkg}$} & Jy/beam & Estimated uncertainty on $S_{bkg}$\\%
48 & \emph{S$_{err}^{noise}$} & Jy/beam & Estimated uncertainty on $S$\\%
49 & \emph{S$_{noise}$} & Jy/beam & Sum of background noise rms extrapolated over number of beams in source\\%
50-51 & (\emph{S$_{tot}$},\emph{S$_{tot,err}$}) & Jy/beam & Sum of source pixel brightness with background subtracted and its uncertainty\\%
52-53 & (\emph{$F$},\emph{$F_{err}$}) & Jy & Source measured flux density (with background subtracted) and its statistical \\
& & & uncertainty\\%
54-55 & (\emph{S$_{tot}^{nested}$},\emph{S$_{tot,err}^{nested}$}) & Jy/beam & Sum of pixel brightness of point-like/compact sources (with background  \\%
& & & subtracted) found inside this source and its uncertainty\\%
56-57 & (\emph{S$_{tot}^{noPS}$},\emph{S$_{tot,err}^{noPS}$}) & Jy/beam & Sum of source pixel brightness (with background and nested compact sources  \\%
& & & subtracted) and its uncertainty\\%
58-59 & (\emph{F$^{noPS}$},\emph{F$_{err}^{noPS}$}) & Jy & Source measured flux density (with background and nested compact sources \\%
& & & subtracted) and its statistical uncertainty\\%
60 & \emph{is\_flux\_reliable} & - & Flux density is likely
                                                reliable considering theoretical
                                                spatial filtering \{0=\texttt{NO}, 1=\texttt{YES}\} \\%
61 & \emph{border} & - &Source is in the tile border \{0=\texttt{NO}, 1=\texttt{YES}\} \\%
62 & \emph{classname} & - & Source classification label in string format \\%
63 & \emph{classid} & - & Source classification id\{-1=\texttt{MULTICLASS}, 0=\texttt{UNCLASSIFIED}, 1=\texttt{LBV/WR}, \\ 
& & & 2=\texttt{GALAXY}, 3=\texttt{PN}, 4=\texttt{SNR}, 6=\texttt{HII}\}\\%
64 & \emph{objname} & - & Matched catalogue object name(s)\\%
\hline                                             
\end{tabular}
\end{table*}

\end{appendix}
\end{document}